\documentstyle[12pt,epsf]{article}

\makeatletter
\@addtoreset{equation}{section} 
\makeatother

\def\be{\begin{equation}}
\def\ee{\end{equation}}
\def\beq{\begin{eqnarray}}
\def\eeq{\end{eqnarray}}
\def\bd{\begin{displaymath}}
\def\ed{\end{displaymath}}

\newcommand{\gapproxeq}{\lower .7ex\hbox{$\;\stackrel{\textstyle >}{\sim}\;$}}
\newcommand{\lapproxeq}{\lower .7ex\hbox{$\;\stackrel{\textstyle <}{\sim}\;$}}

\begin{document}
\setlength{\unitlength}{1mm}

{\hfill astro-ph/0212518}

{\hfill DSF 26/02}\vspace*{1cm}

\begin{center}
{\Large \bf Coupled quintessence and the coincidence problem}
\end{center}

\bigskip\bigskip

\begin{center}
{\bf G. Mangano}, {\bf G. Miele}, and {\bf V. Pettorino}
\end{center}

\vspace{.5cm}

\noindent {\it Dipartimento di Fisica, Universit\'{a} di Napoli
"Federico II", and INFN, Sezione di Napoli, Complesso
Universitario di Monte
S. Angelo, Via Cintia, I-80126 Naples, Italy}\\
\bigskip\bigskip\bigskip

\begin{abstract}
We consider a model of interacting cosmological constant/quintessence,
where dark matter and dark energy behave as, respectively, two coexisting
phases of a fluid, a thermally excited Bose component and a condensate,
respectively. In a simple phenomenological model for the dark components
interaction we find that their energy density evolution is strongly coupled
during the universe evolution. This feature provides a possible way out for
the coincidence problem affecting many quintessence models.
\end{abstract}

\vspace{.5cm}

\begin{center}
PACS 98.80.-k, 98.80.Cq

\end{center}

The extensive astrophysical observations of Supernovae (SN) Ia provide a
powerful tool to single out the main cosmological energy density
contributions at redshift $z \sim 1$. The evidence for a presently
accelerated cosmic expansion \cite{Perlmutter,Riess} suggests the presence
of a new dominant energy source whose equation of state $p=\omega\rho$ is
compatible with a cosmological constant. The analysis of the recent
experimental data favors in fact the range $-1\leq \omega
\leq -0.4$ \cite{Perlmutter,Turner&White,Wangetal}. In particular
for a pure cosmological
constant the favored energy density, in unit of the critical density, would
be of the order of $\Omega_\Lambda
\simeq 0.7$. Unfortunately, this value would be several
orders of magnitude smaller than what can be predicted by our present
knowledge of fundamental interactions. Moreover, it is also quite unnatural
that $\Omega_\Lambda$ is just of the same order of the cold dark matter
term $\Omega_m
\simeq 0.3$ today, while it has been negligible for, say, red-shifts larger than 1.
This is a severe fine tuning problem, usually referred to as the {\it
coincidence problem}. \\A much more appealing scenario is to look at this
additional term as due to a scalar field $\phi$ whose dynamics makes it
negligible at high redshift and dominating the cosmic energy density at
$z\leq 1$ \cite{Wetterich,Ratra&Peebles,Frieman&al,Caldwell&al}. However
this wide class of model is not free from fine tuning problem, due to the
arbitrariness in the choice of the scalar field potential, but it has
nevertheless more appealing features. This is for example the case of {\it
tracker fields} scenarios, where the scalar dynamics shows a
pseudo-attractor solution, valid for a wide range of initial conditions and
providing values of $\omega_\phi$ and $\Omega_\phi$
\cite{Wetterich,Ratra&Peebles} in good agreement with SN Ia observations.

Recently interacting quintessence scenarios have been extensively studied,
considering a scalar field coupled with gravity, ordinary matter and dark
matter \cite{Carroll,Damour,Uzan,Chiba,Chen,Baccigalupi}. In particular,
the recent data on CMBR anisotropy have provided a chance to constrain
quintessence models
\cite{Amendola,Netterfield,Lee,Halverson,Corasaniti,Bean,Tocchini,Amendola02},
though the high degree of arbitrariness on the scalar dynamics makes these
constraints very model dependent. In this letter we assume a more
phenomenological scenario, in which the quintessence and dark matter
components are looked upon as two different components of the same fluid, a
condensate contribution and a thermally Bose distributed "gas", with energy
densities $\rho_\Lambda$ and $\rho_m$, respectively. We start considering
the case where the condensate energy term corresponds to a pure
cosmological constant, $\omega=-1$. As it will be clear in the following,
the main features of the model still hold for a more general equation of
state with $\omega<0$. The gas component $\rho_m$ is assumed to be cold
matter, with vanishing pressure. Finally we consider for simplicity a
spatially flat universe, $\Omega_{k}=0$.

We parametrize the interaction in terms of powers of the two energy
densities $\rho_\Lambda$ and $\rho_m$. The time evolution of radiation
energy density, $\rho_R$, consensate and dark matter are then dictated by
the following equations
\begin{eqnarray}
\dot{\rho }_{r} & = & -4 \, \frac{\dot{a}}{a} \, \rho _{r} \,\,\, ,\label{eq1} \\
 &  & \nonumber \\
\dot{\rho }_{m} & = & -3 \, \frac{\dot{a}}{a} \, \rho _{m}- k \, M^{5-4\alpha -4\beta } \,
\rho ^{\alpha }_{m} \, \rho^{\beta }_{\Lambda } \,\,\, ,\label{eq2} \\
 &  & \nonumber \\
\dot{\rho }_{\Lambda } & = & k \, M^{5-4\alpha -4\beta }
 \, \rho ^{\alpha }_{m} \, \rho ^{\beta }_{\Lambda }
\, \, \, ,\label{eq3}
\end{eqnarray}
where $k$ is a dimensionless coupling constant and $M$ a suitable mass
parameter. In terms of the redshift $z$ equations (\ref{eq1})-(\ref{eq3})
can be conveniently rewritten as
\begin{eqnarray}
\frac{d\tilde{\rho }_{r}(z)}{dz} & = & \frac{4}{1+z} \, \tilde{\rho }_{r}(z)\,\,\, ,
\label{eq1vsz} \\
 &  & \nonumber \\
\frac{d\tilde{\rho }_{m}(z)}{dz} & = & \frac{1}{1+z}\, \left( 3 \,\tilde{\rho }_{m}(z)+
\lambda \, \frac{\tilde{\rho }^{\alpha }_{m}(z)\, \tilde{\rho }^{\beta }_{\Lambda }(z)}{\sqrt{\tilde{\rho }_{r}(z)
+\tilde{\rho }_{m}(z)+\tilde{\rho }_{\Lambda }(z)}}\right)
\,\,\, ,\label{eq2vsz}
\\
 &  & \nonumber \\
\frac{d\tilde{\rho }_{\Lambda }(z)}{dz} & = & - \frac{\lambda}{1+z} \, \frac{\tilde{\rho }^{\alpha }_{m}(z)\,
\tilde{\rho }^{\beta }_{\Lambda }(z)}{\sqrt{\tilde{\rho }_{r}(z)+\tilde{\rho }_{m}(z)+\tilde{\rho }_{\Lambda }(z)}}
\,\,\, ,\label{eq3vsz}
\end{eqnarray}
where $\tilde{\rho }_{i}\equiv \rho _{i}/\rho _{c_{0}}$, with $\rho
_{c_{0}}=3 H^2_0 M_{Pl}^2/(8 \pi)= h^2 \, 0.81 {\cdot} 10^{-46}\, GeV^4$
the present critical energy density, and
\begin{equation}
\lambda \equiv k \, \sqrt{\frac{3}{8 \pi}} \,
\left(\frac{M_{Pl}}{M}\right) \, \left(\frac{\rho _{c_{0}}}
{M^4}\right)^{\alpha+\beta-3/2}\,\,\,. \label{deflambda}
\end{equation}
By defining $x\equiv \log
_{10}\left( 1+z\right)$ we then get
\begin{eqnarray}
\frac{d\tilde{\rho }_{r}(x)}{dx} & = & 4 \, \log(10) \, \tilde{\rho }_{r}(x)\,\,\, ,\label{eq1vsx} \\
 &  & \nonumber \\
\frac{d\tilde{\rho }_{m}(x)}{dx} & = & \log(10)\, \left( 3 \,
\tilde{\rho }_{m}(x)+\lambda \, \frac{\tilde{\rho }^{\alpha
}_{m}(x) \, \tilde{\rho }^{\beta }_{\Lambda
}(x)}{\sqrt{\tilde{\rho }_{r}(x) +\tilde{\rho }_{m}(x)+\tilde{\rho
}_{\Lambda }(x)}}\right) \,\,\, ,\label{eq2vsx}
\\
 &  & \nonumber \\
\frac{d\tilde{\rho }_{\Lambda }(x)}{dx} & = & -\lambda \, \log(10)
\,\frac{\tilde{\rho }^{\alpha }_{m}(x) \, \tilde{\rho }^{\beta
}_{\Lambda }(x)}{\sqrt{\tilde{\rho }_{r}(x)+\tilde{\rho }_{m}(x)+
\tilde{\rho }_{\Lambda }(x)}} \,\,\, .\label{eq3vsx}
\end{eqnarray}
Note that for the parameter $\lambda < 0$, the interaction term on the
r.h.s. of (\ref{eq2vsx}) and (\ref{eq3vsx}) describes a continuous transfer
of energy from the condensate to the ordinary non-relativistic particles.
The corresponding scenario is a universe filled by both the condensate and
a normal gas of dark particles, but with an effective mass of the
condensate larger than the mass of thermal excitation in the gas phase.
Direct decay of the condensate in dark matter particles is expected to be
proportional to $\rho_\Lambda$ only, while stimulated decay mechanism is of
course proportional to number density of dark matter particles too. The
natural expectation in this case is an interaction term with
$\alpha=\beta=1$. In the following we will keep varying these parameters,
to study the general features of the model as function of $\alpha$ and
$\beta$ and eventually study in more details the case whereboth are set
equal to unity.

We will be mainly interested to the case $\lambda<0$. The case $\lambda>0$,
which will be briefly treated in the following, also represents a possible
framework. In this case, as the temperature lowers, a fraction of the
energy stored in the gas of dark particles flows in the condensate, as
during a temperature driven phase transition. Unfortunately this model
suffers of the very same coincidence problem of many other quintessence
scenario and thus it is much less appealing.

The system (\ref{eq1vsx})-(\ref{eq3vsx}) can be numerically solved, as a
function of $x$, by fixing the values of $\tilde{\rho }_{r }$, $\tilde{\rho
}_{m}$ and $\tilde{\rho }_{\Lambda }$ at present ($x_i=0$), and then
following the values of the energy densities back in time. In particular,
to describe the features of this coupled system, we have chosen as
reference values $\tilde{\rho }_{m }(0)=\Omega_m=0.3$, $\tilde{\rho
}_{\Lambda }(0)=\Omega_\Lambda=0.7$ and the {\it equivalence} red-shift
$z_{eq}
\simeq 3100$. This implies
\begin{equation}
\frac{\rho _{r}(z)}{\rho _{m}(z)}=\frac{\rho _{r}(0)}{\rho _{m}(0)}(1+z)
\Longrightarrow
\frac{\rho _{r}(0)}{\rho _{m}(0)}=\frac{1}{1+z_{eq}}\,\,\, .
\end{equation}
It may seem that fixing the values of the energy densities parameters today
is already a fine-tuning condition imposed to the evolution. This of course
would be true if no interactions would be considered in the model. We will
show in fact that, since $\rho_\Lambda$ and $\rho_m$ evolution are coupled,
the fact that their ratio $\Omega_\Lambda/\Omega_m$ is of order one today
represents a much less ad hoc condition for a wide range of the parameters
$\alpha$, $\beta$ and $\lambda$. The fact that $\Omega_\Lambda
\sim \Omega_m$ represent the main contribution to the energy density today
is ultimately related to the fact that the interaction term keeps their
evolution quite close throughout the universe expansion. We can say that
the only remaining fine tuned condition is given by the value of the
equivalence redshift, where radiation start becoming negligible.

In order to study the role of the free parameter $\alpha$, $\beta$,
$\gamma$ on the evolution of the energy densities, we start fixing for
example $\lambda =
- 10^{-3}$ and choose $\alpha=1$ and $\beta=0$. In Figure
\ref{densita` alfa1beta0} are reported the energy densities for radiation
$\tilde{\rho}
_{r}$, matter $\tilde{\rho} _{m}$ and condensate $\tilde{\rho} _{\Lambda}$,
as a function of $x$. For these values of free parameters, the interaction
is due to $\tilde{\rho}_{m}$ only, but differently from the noninteraction
case, the energy density of condensate here evolves in time. The evolution
is of course enhanced  for larger values of $\alpha$, but for $x \lapproxeq
2$ the condensate energy density $\tilde{\rho}_{\Lambda}$ reaches a
constant value which, at present, provides an effective cosmological
constant. Thus for $\beta=0$ the model predicts scenarios very similar to
other quintessence models and no new features overcoming the coincidence
problem appear.

\begin{figure}
\begin{center}
\begin{tabular}{|cc|}
\hline
& \\
\epsfxsize=7cm \epsfysize=4.5cm \epsffile{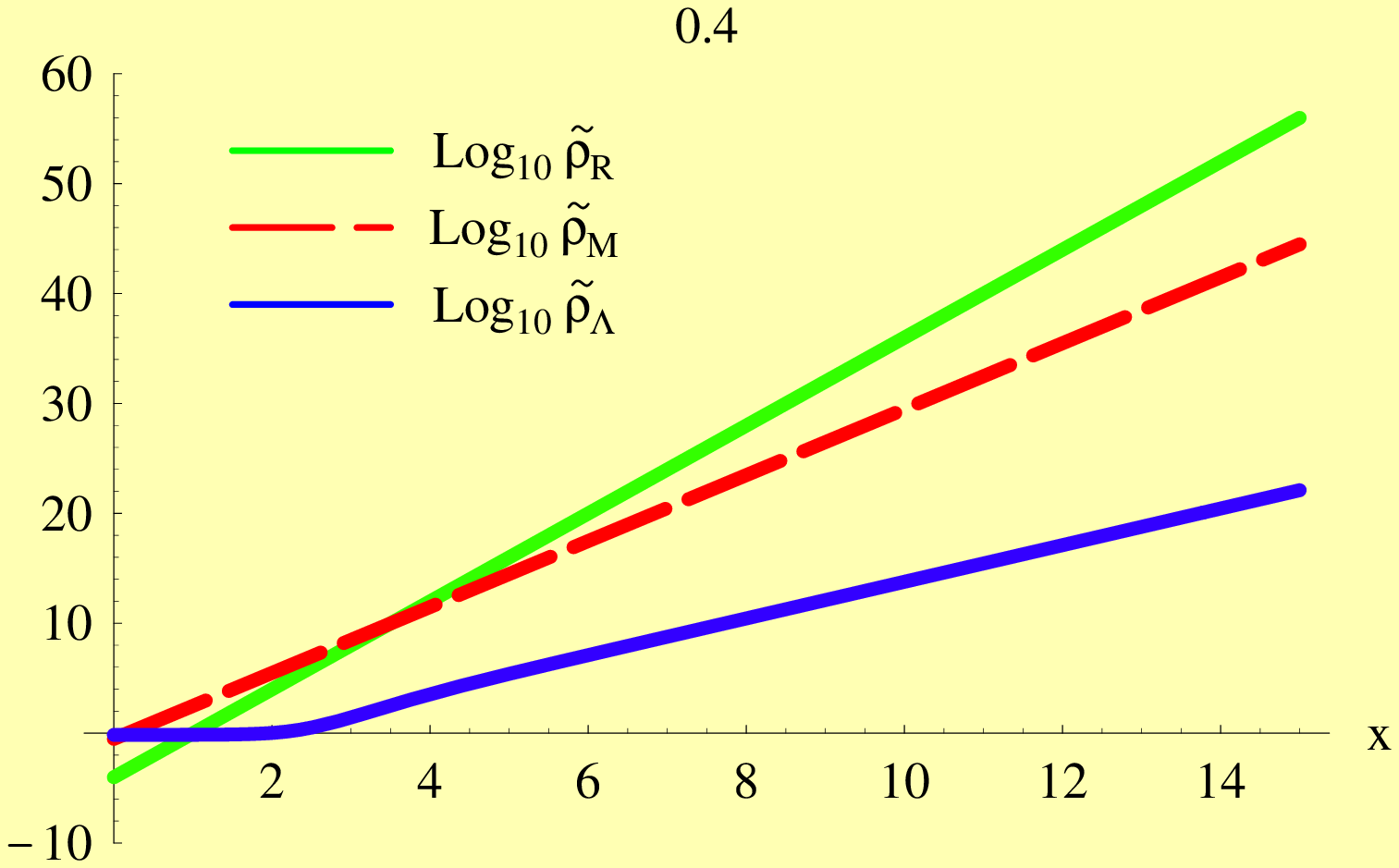} &
\epsfxsize=7cm \epsfysize=4.5cm \epsffile{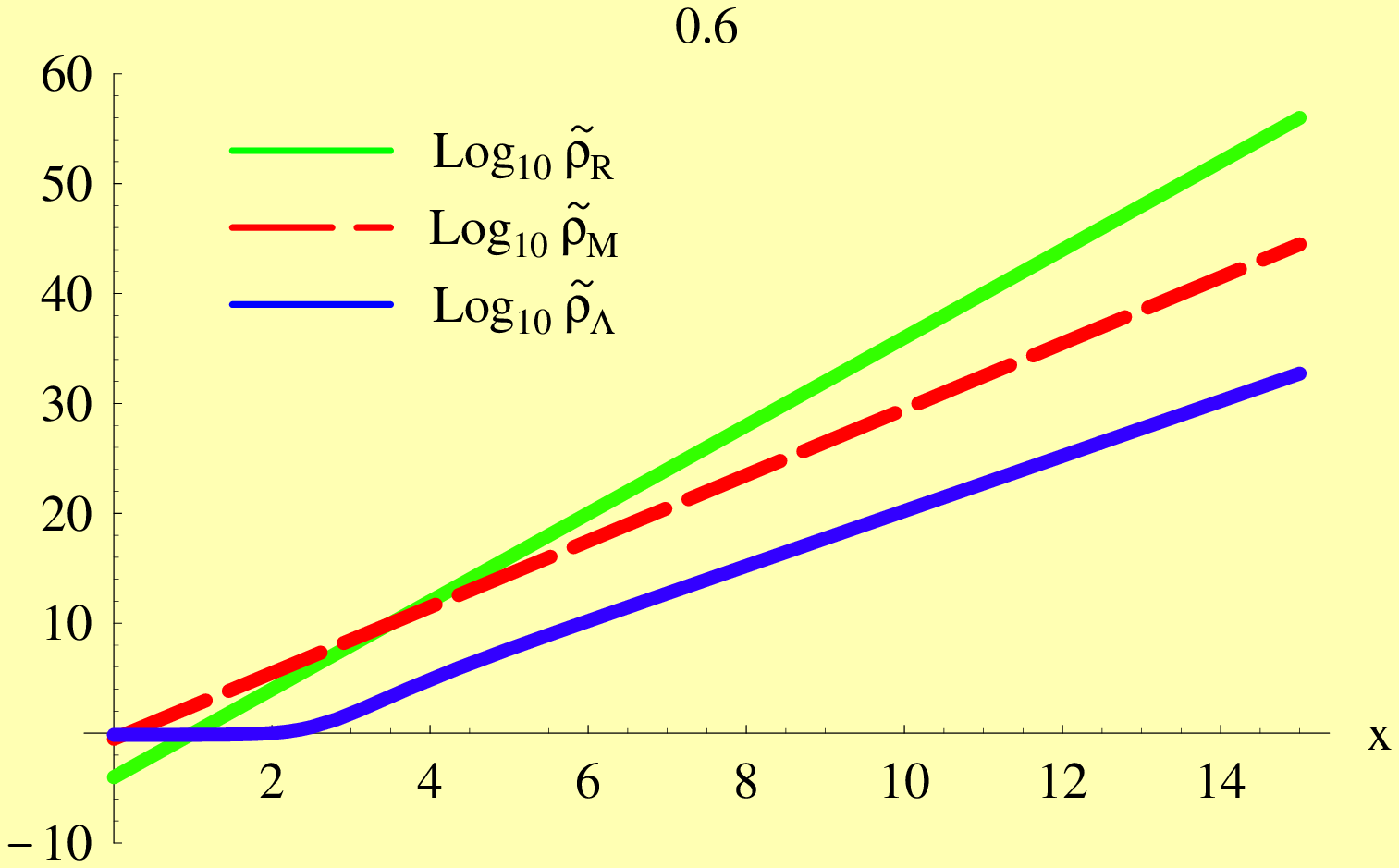}
\\
\epsfxsize=7cm \epsfysize=4.5cm \epsffile{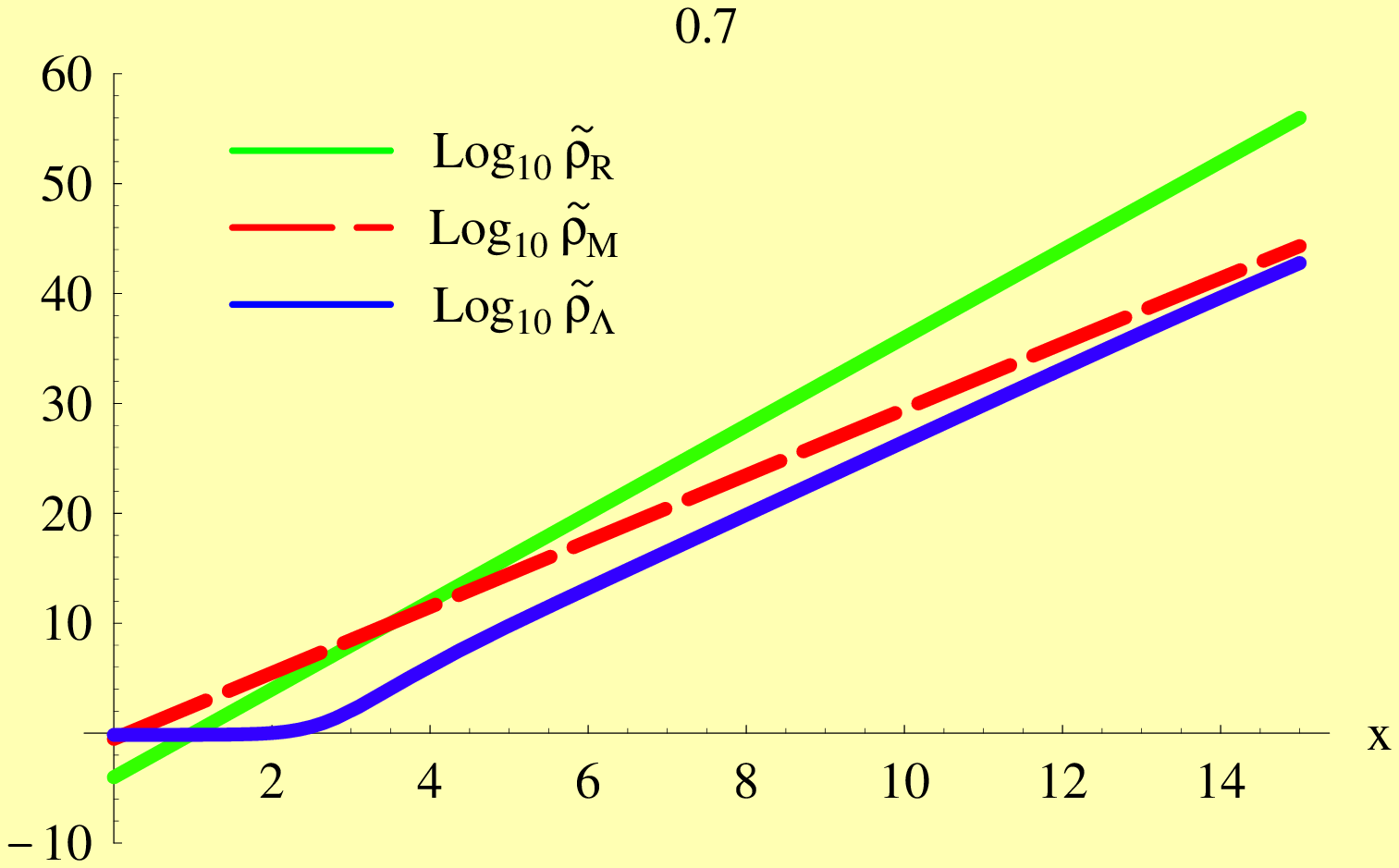} &
\epsfxsize=7cm \epsfysize=4.5cm \epsffile{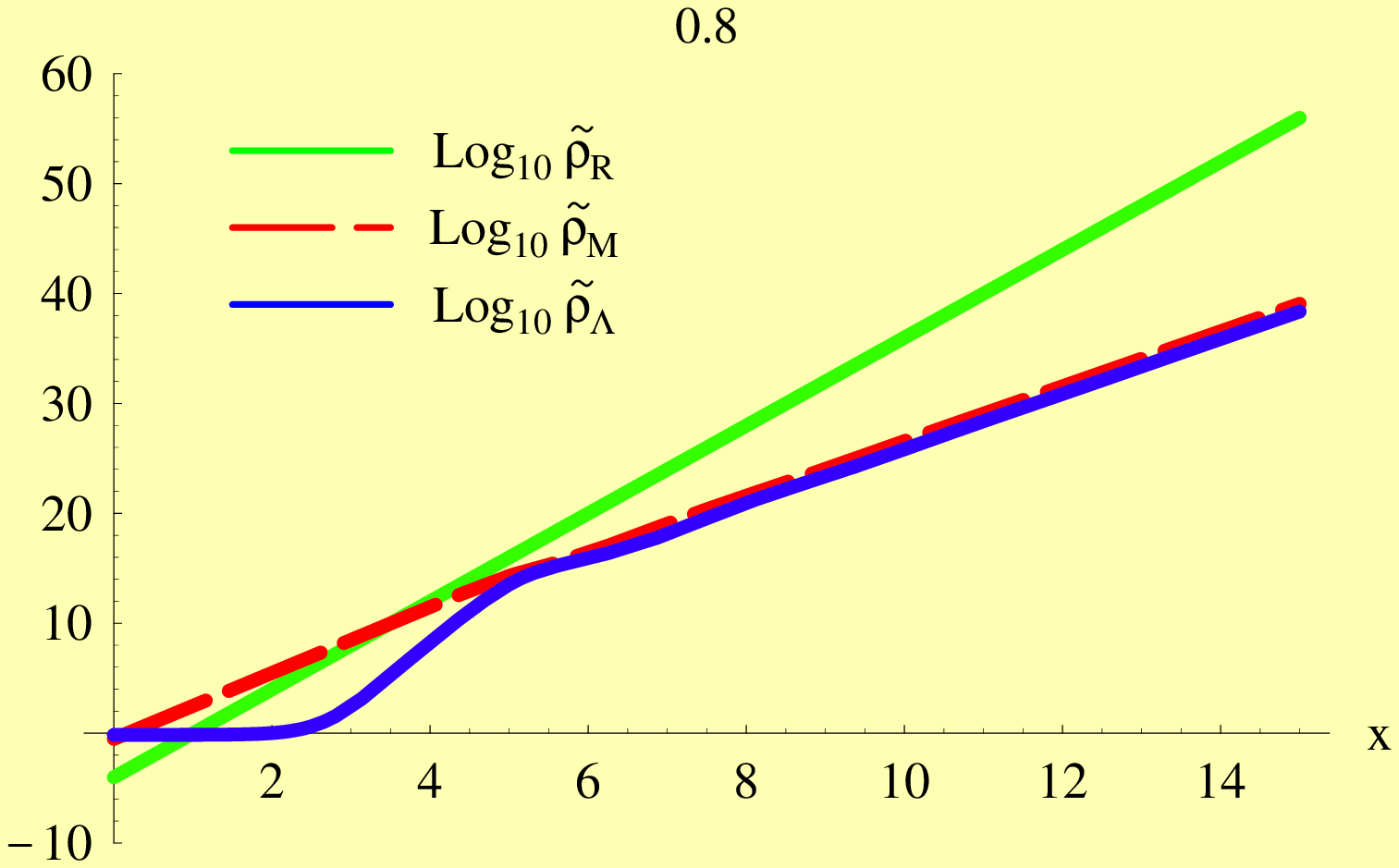}
\\
\epsfxsize=7cm \epsfysize=4.5cm \epsffile{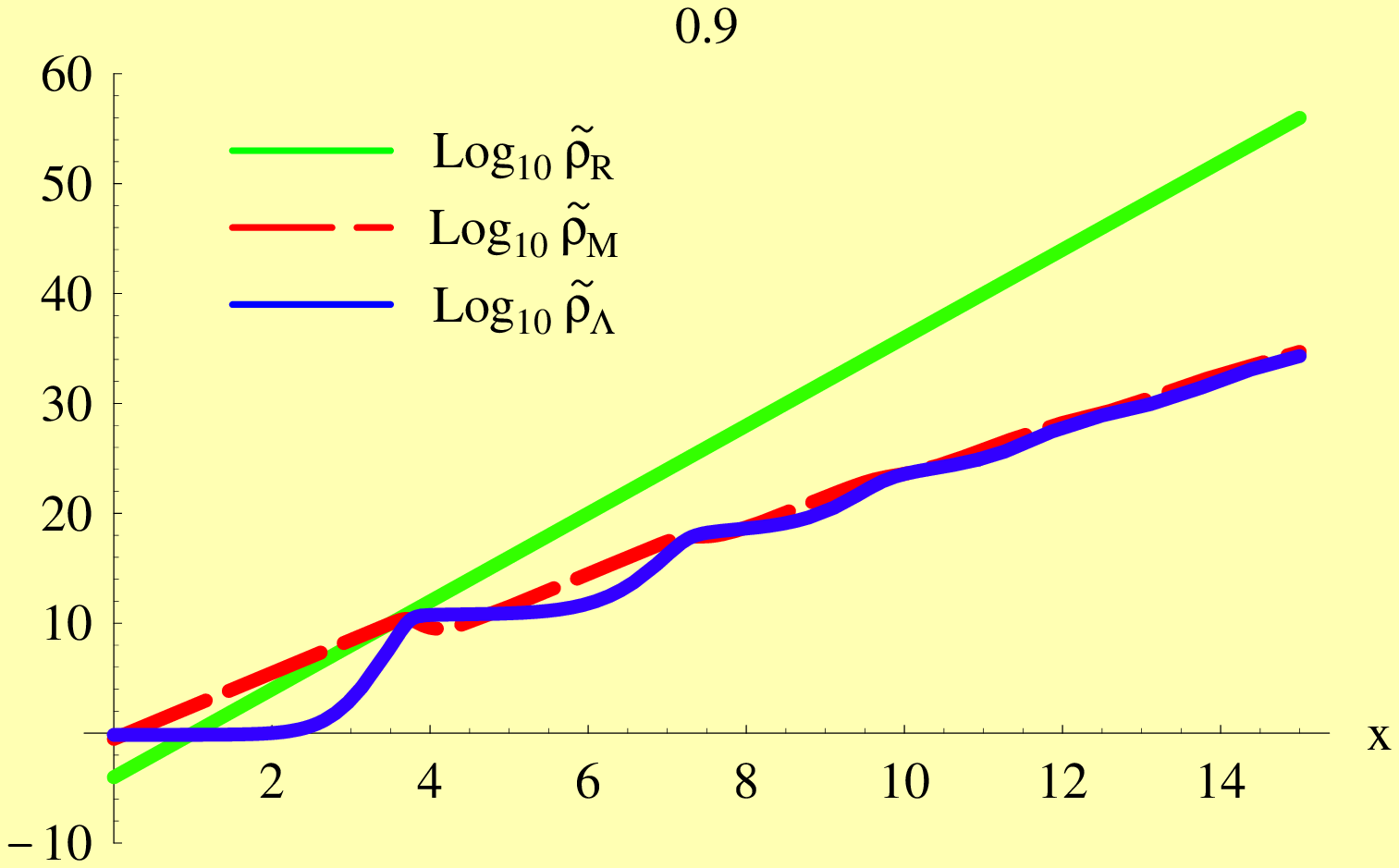} &
\epsfxsize=7cm \epsfysize=4.5cm\epsffile{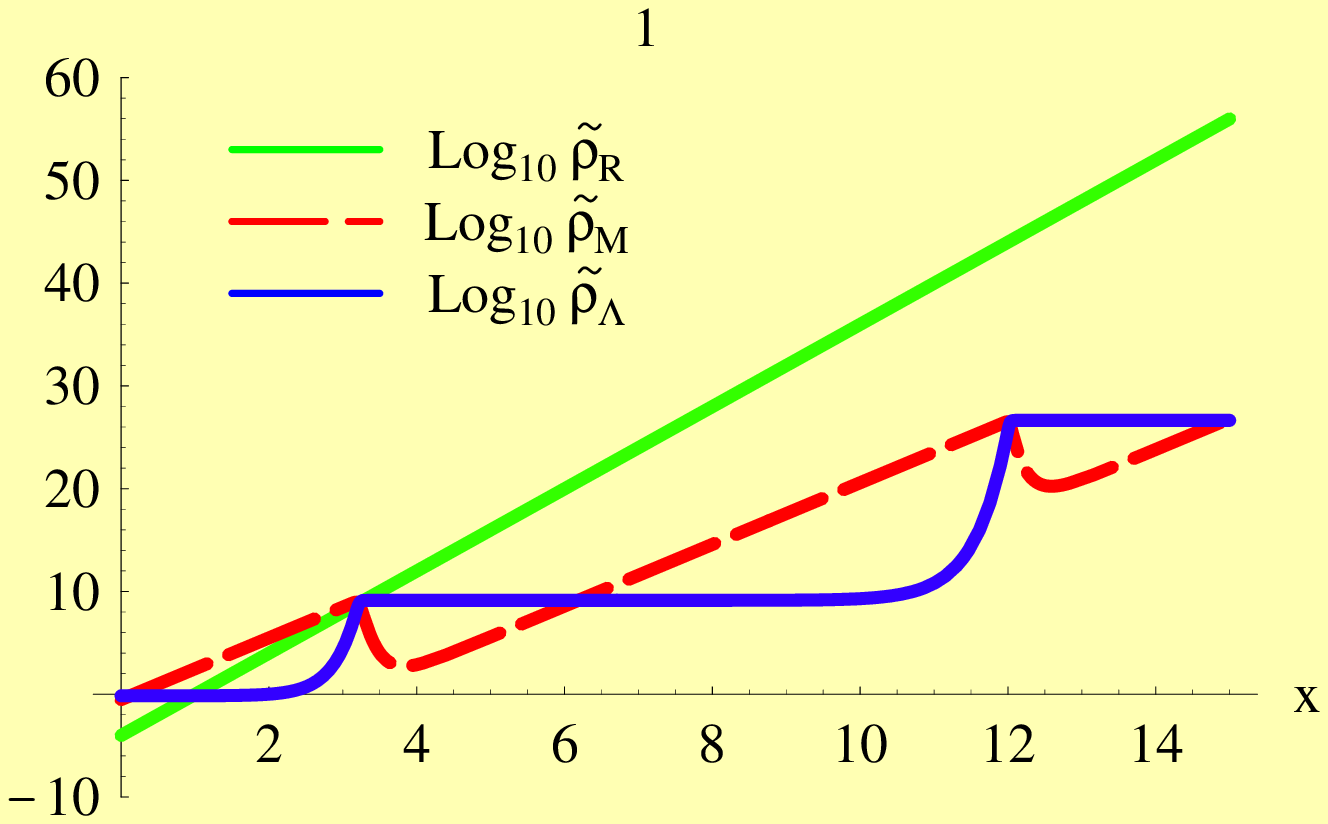}\\
\hline
\end{tabular}
\end{center}
\caption{Energy densities in unit of $\rho _{c_{0}}$ for
$\alpha=1$ and $\lambda=-0.001$ and varying $\beta$; the corresponding
value of $\beta$ is reported on the top of each graph.}
\label{variabeta}
\end{figure}

The behavior of the system completely changes if one takes non vanishing
values for $\beta$. In this case, by increasing the value of $\beta$, the
values of dark energy $\tilde{\rho}_{\Lambda}$ and dark matter
$\tilde{\rho}_{m}$ are more and more connected. As it is clear from Figure
\ref{variabeta}, where the energy densities are plotted versus $x$ for
different values of $\beta$ and for $\alpha=1$ and $\lambda=-0.001$, due to
the interaction term Eq.s (\ref{eq2vsx}) and (\ref{eq3vsx}),
$\tilde{\rho}_{m}$ decreases for large redshift with the increasing of
$\beta$, whereas $\tilde{\rho}_{\Lambda}$ still remains with larger values.
As a consequence their evolution remain closer. The two quantities start to
intersect each other for values of $\beta$ larger than $0.8$. This is a
relevant new feature of the model which solves the problem of coincidence,
simply because the situation where the two densities
$\tilde{\rho}_{\Lambda}$ and $\tilde{\rho}_{m}$ are of the same order of
magnitude it is not a peculiarity of the present epoch but occurred many
times in past.

\begin{figure}
\begin{center}
\epsfxsize=13cm \epsfysize=8cm
\epsffile{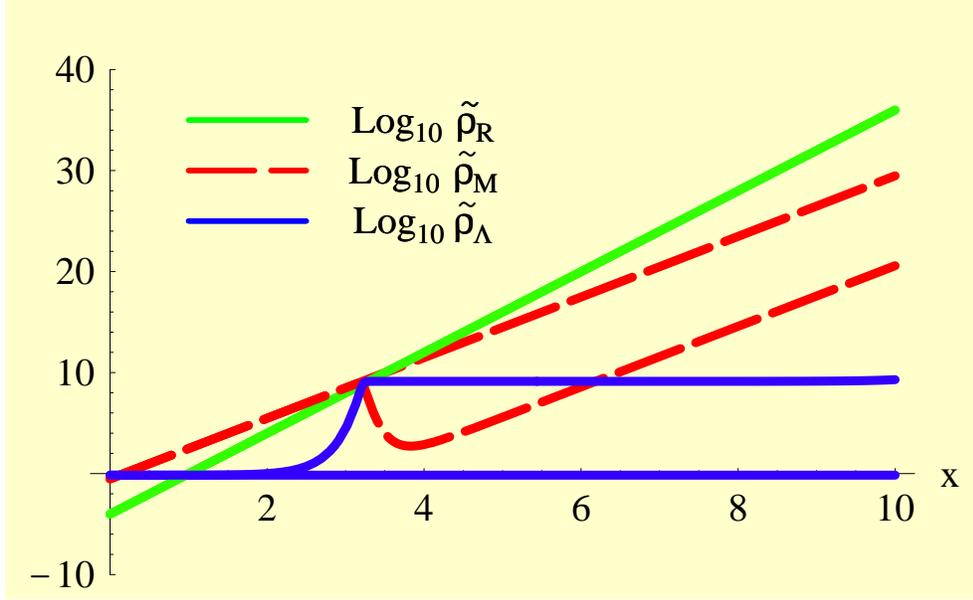}
\end{center}
\caption{Energy densities in unit of $\rho _{c_{0}}$ for $\lambda
=0$, $-0.001$ with $\alpha =1$ and $\beta =1$.} \label{tuning}
\end{figure}

In Figure \ref{tuning} we show the plots corresponding to the
non-interacting case $\lambda=0$ and the evolution for $\lambda=-0.001$ and
$0$ for $\alpha,\,\beta
=1$ . In this case for $0 < x < 2 $ the two graphs exactly
overlap, since in Eq. (\ref{eq2vsx}) the first term dominates whereas in
Eq. (\ref{eq3vsx}) the l.h.s. is almost vanishing. This feature of the
model is extremely important because provides a proper behavior of dark
matter and energy for redshift smaller than $z_{eq}$ which does not affect
large scale structures formation.

The behavior shown by the energy densities plots for $\lambda
=-0.001$ and $\beta=1$, reported in Figures \ref{variabeta} and
\ref{tuning} can be easily understood. By increasing the redshift
$x$ the adimensionalized Hubble parameter $
\tilde{H}=\sqrt{\tilde{\rho }_{r}+\tilde{\rho }_{m}+ \tilde{\rho
}_{\Lambda }}$ at the denominator of Eq.s (\ref{eq2vsx}) and (\ref{eq3vsx})
also increases, but less than the product of energy densities at the
numerator does. Thus for large values of $\tilde{\rho}_{m}$ the interaction
term in Eq. (\ref{eq3vsz}) starts to be effective, consequently
$\tilde{\rho }_{\Lambda }$ increases hence enhancing the interaction term.
This occurs till $\tilde{\rho }_{\Lambda }$ reaches the dark matter
contribution (at $x \sim 3$ for this choice of parameters). From this value
on $\tilde{\rho }_{m}$ feels the presence of $\tilde{\rho }_{\Lambda }$ and
starts to decrease whereas the dark energy remains constant. This behavior
ends when the first term in the r.h.s of (\ref{eq2vsx}) becomes the
dominant one and thus $\tilde{\rho }_{m}$ starts to increase again and so
the cycle repeats.

\begin{figure}
\begin{center}
\begin{tabular}{|cc|}
\hline
& \\
\epsfxsize=7cm \epsfysize=4.5cm \epsffile{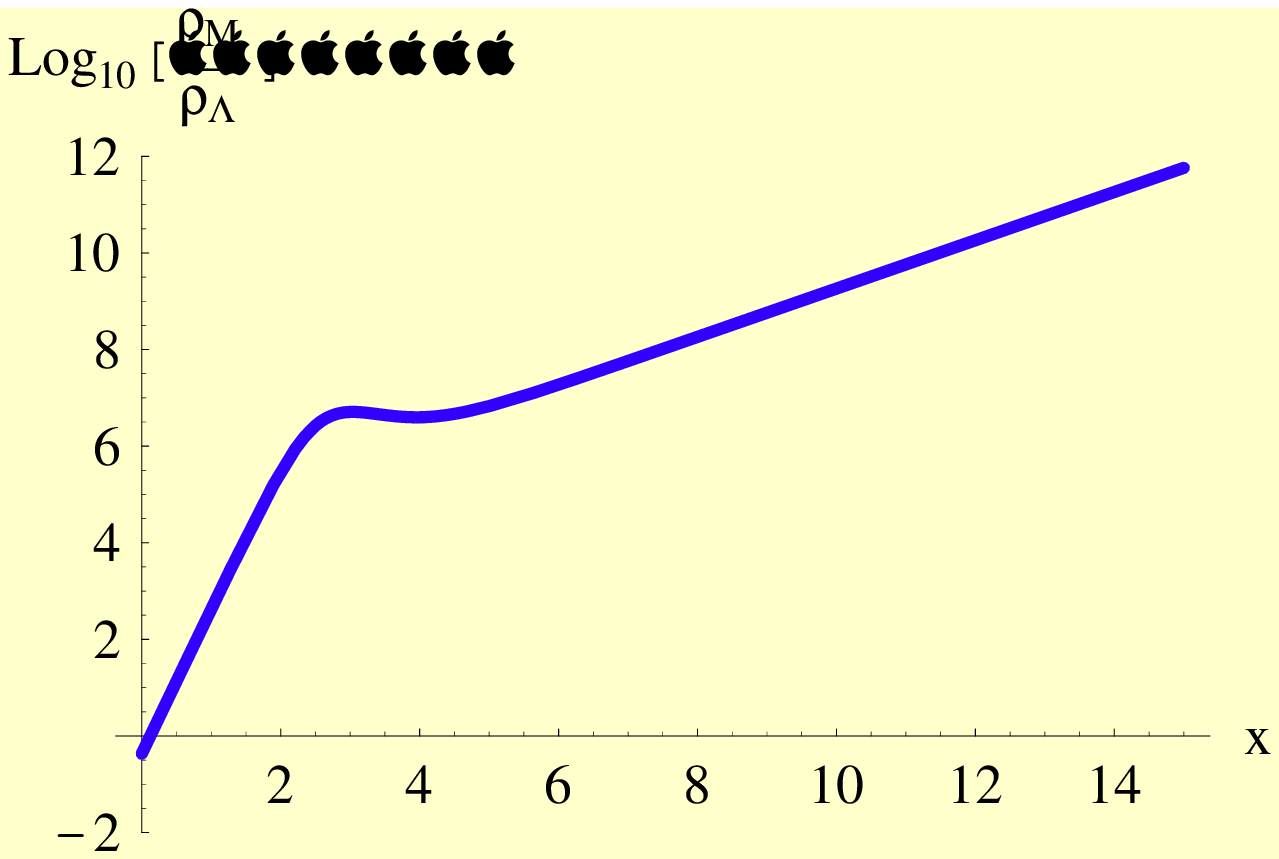} &
\epsfxsize=7cm \epsfysize=4.5cm \epsffile{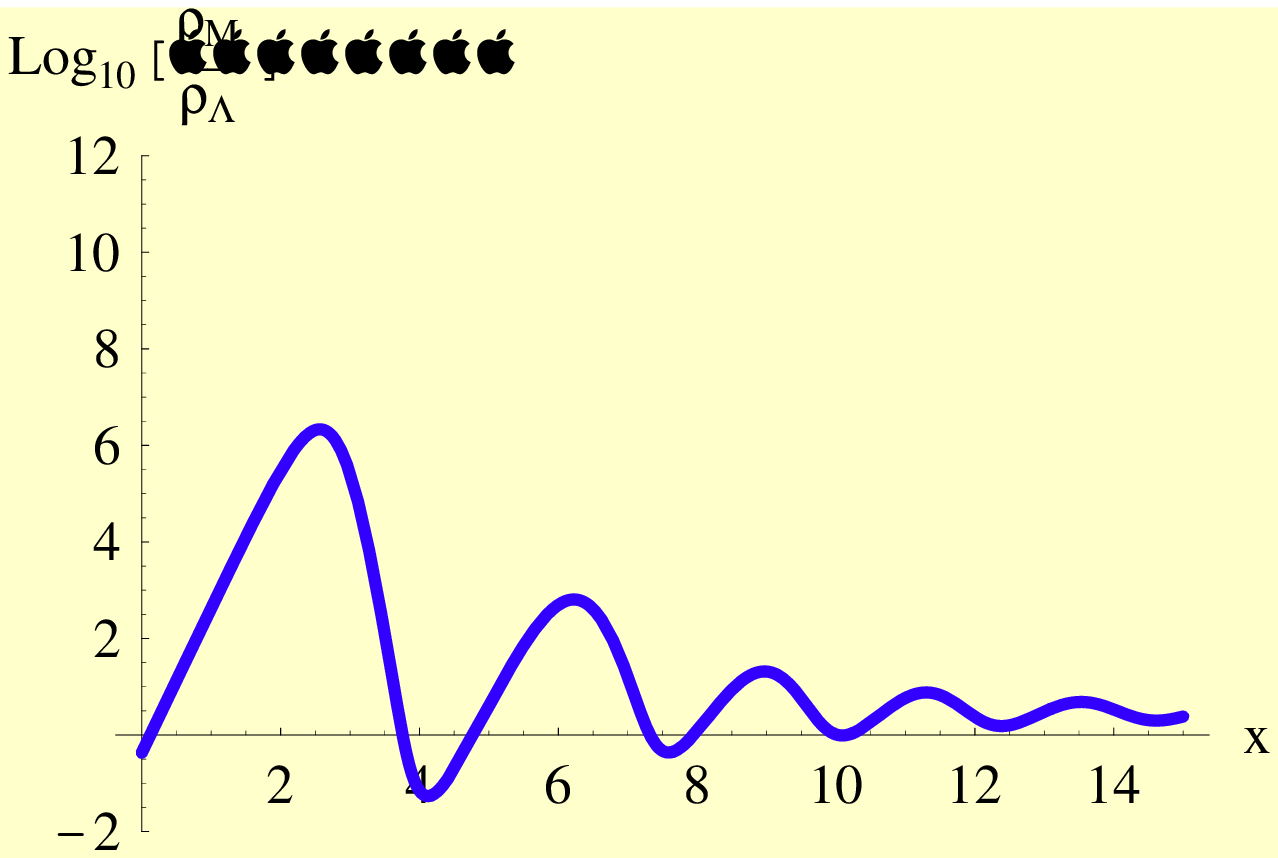}
\\
(a) & (b) \\
\hline
\end{tabular}
\end{center}
\caption{The $\log_{10}\left[\tilde{\rho}_m
/\tilde{\rho}_\Lambda\right]$ versus $x$ for the two values
$\beta=0.6$ (Fig.(a)) and $\beta=0.9$ (Fig.(b)). The values of $\alpha$ and
$\lambda$ are the same of Figure \ref{variabeta}.}
\label{rapvariabeta}
\end{figure}

In Figures \ref{rapvariabeta}(a) and \ref{rapvariabeta}(b) we
report the quantity $\log_{10}(\tilde{\rho}_m
/\tilde{\rho}_\Lambda)$ versus $x$ for $\beta=0.6$ and
$\beta=0.9$, respectively. The other parameters are fixed to be $\alpha=1$
and $\lambda-0.001$ as for the plots of Figure \ref{variabeta}. As it is
clear from Figures \ref{rapvariabeta} for $\beta=0.6$ the interaction term
is too small to produce sizeable effects. This situation drastically
changes for $\beta=1$ where the ratio oscillates with peaks and dips
corresponding to the cycles of the Figure \ref{variabeta}. Actually the
value $\beta \sim 1$ is the value which most naturally fit the stimulated
decay scenario previously outlined.

\begin{figure}
\begin{center}
\begin{tabular}{|cc|}
\hline
& \\
\epsfxsize=7cm \epsfysize=4.5cm \epsffile{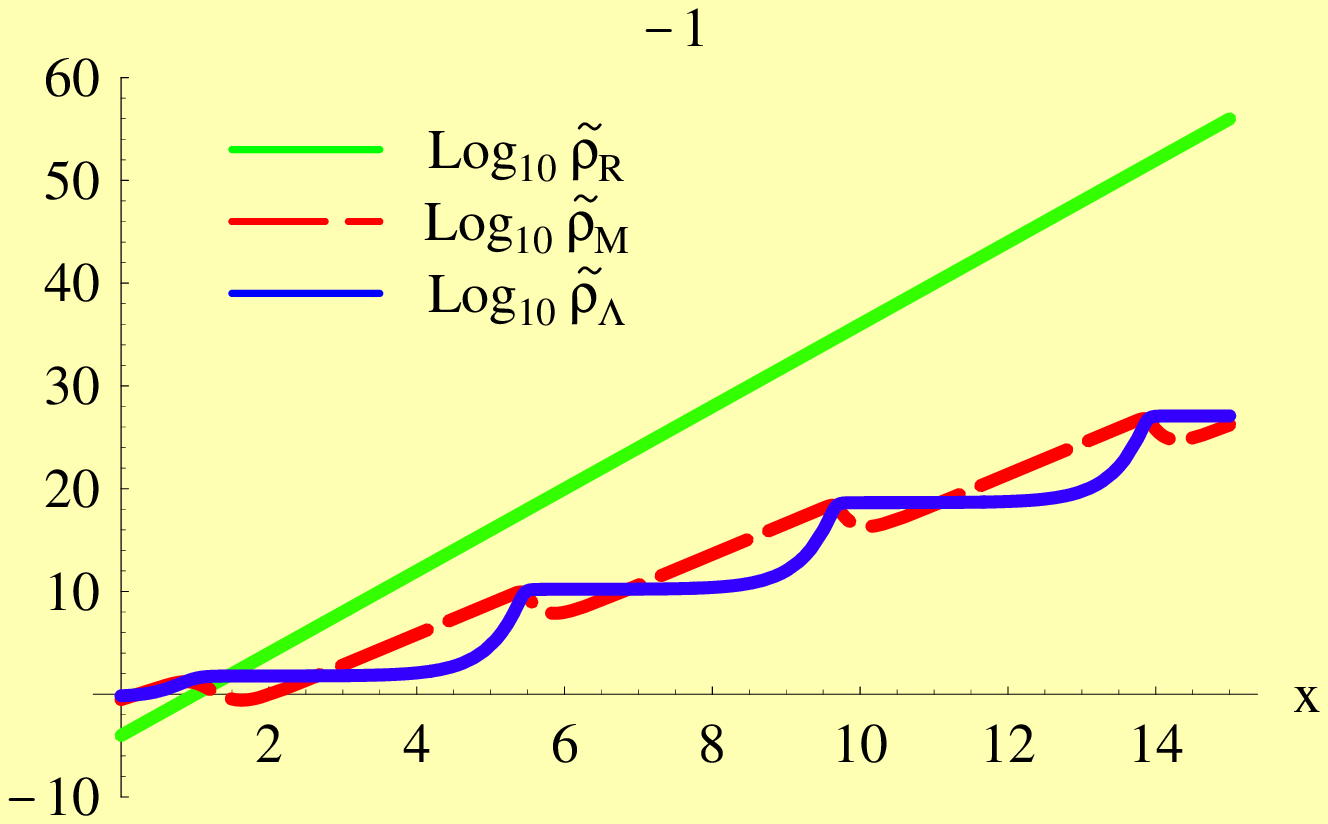} &
\epsfxsize=7cm \epsfysize=4.5cm \epsffile{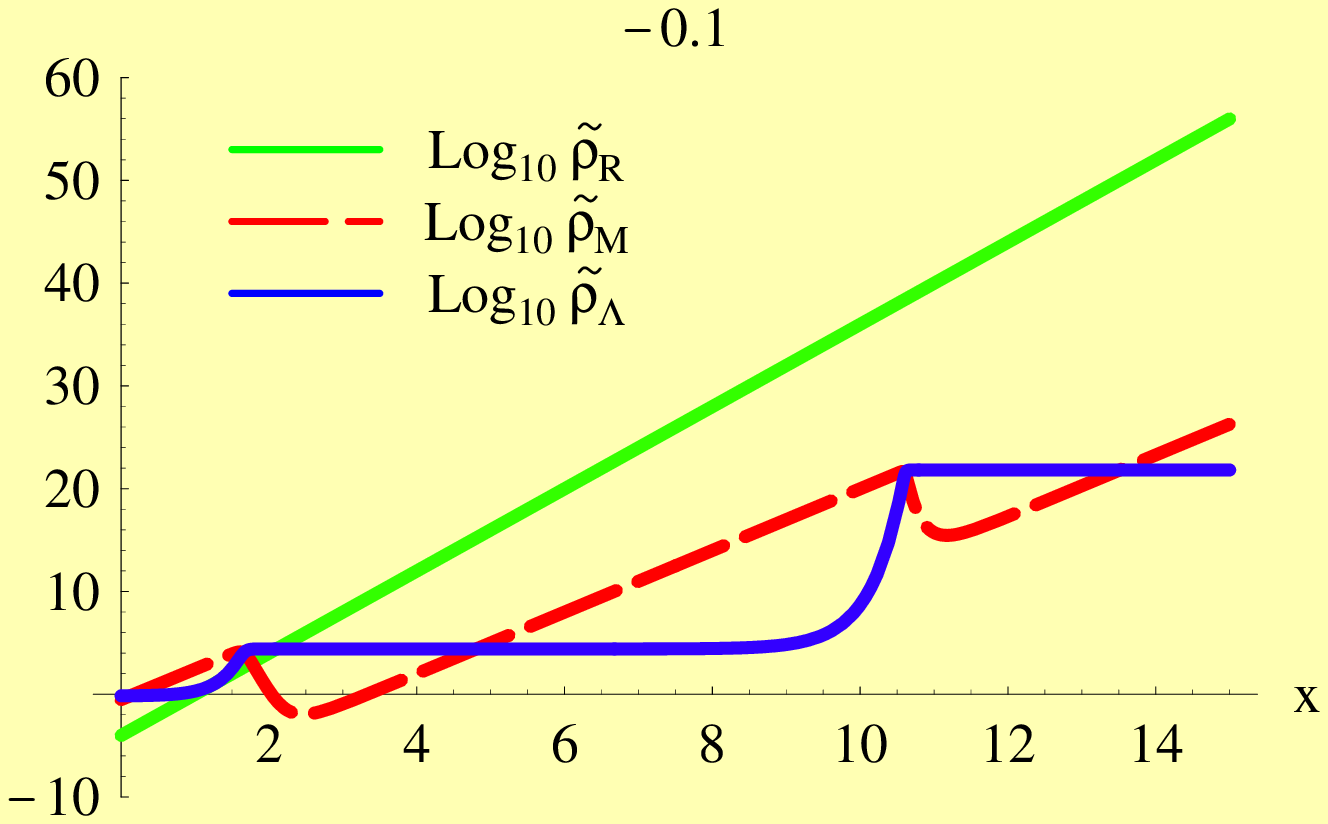}
\\
\epsfxsize=7cm \epsfysize=4.5cm \epsffile{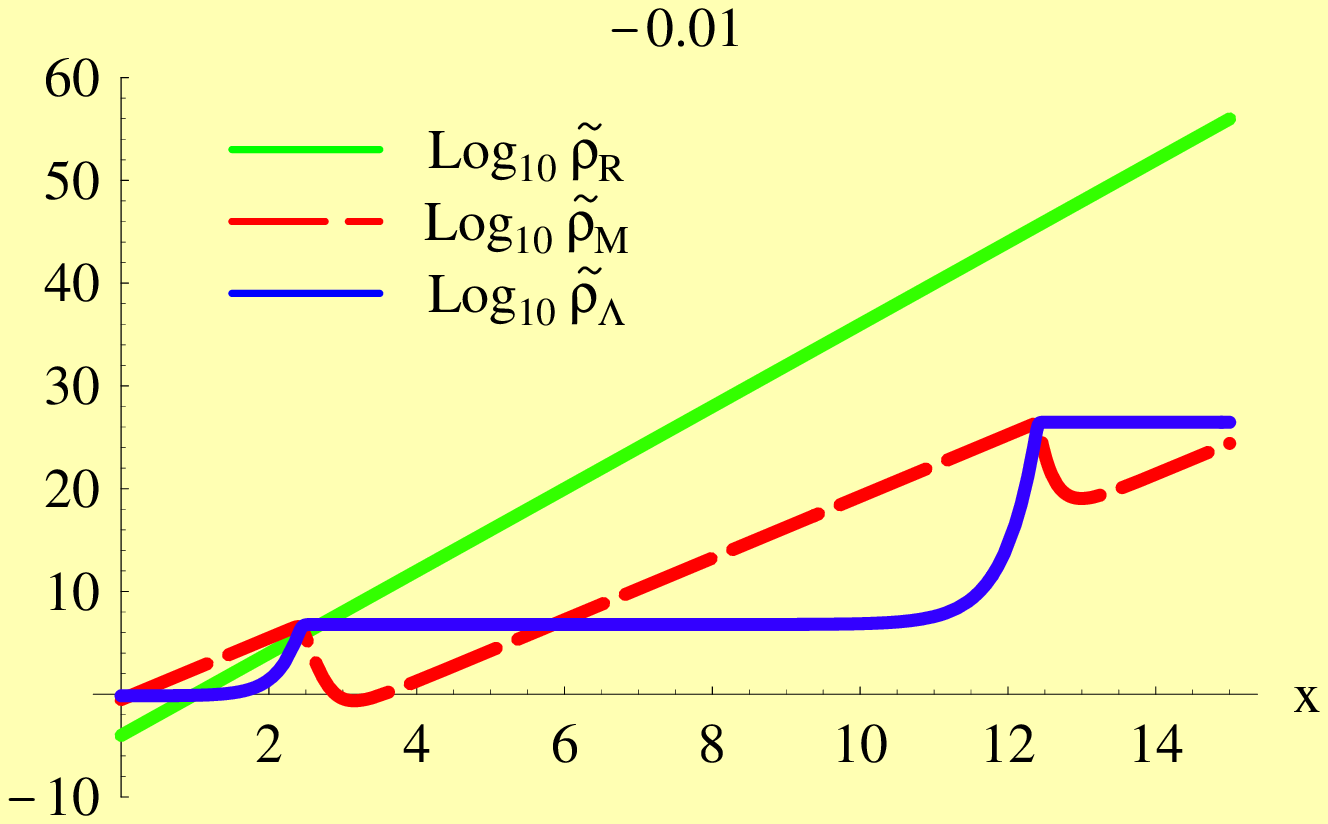} &
\epsfxsize=7cm \epsfysize=4.5cm \epsffile{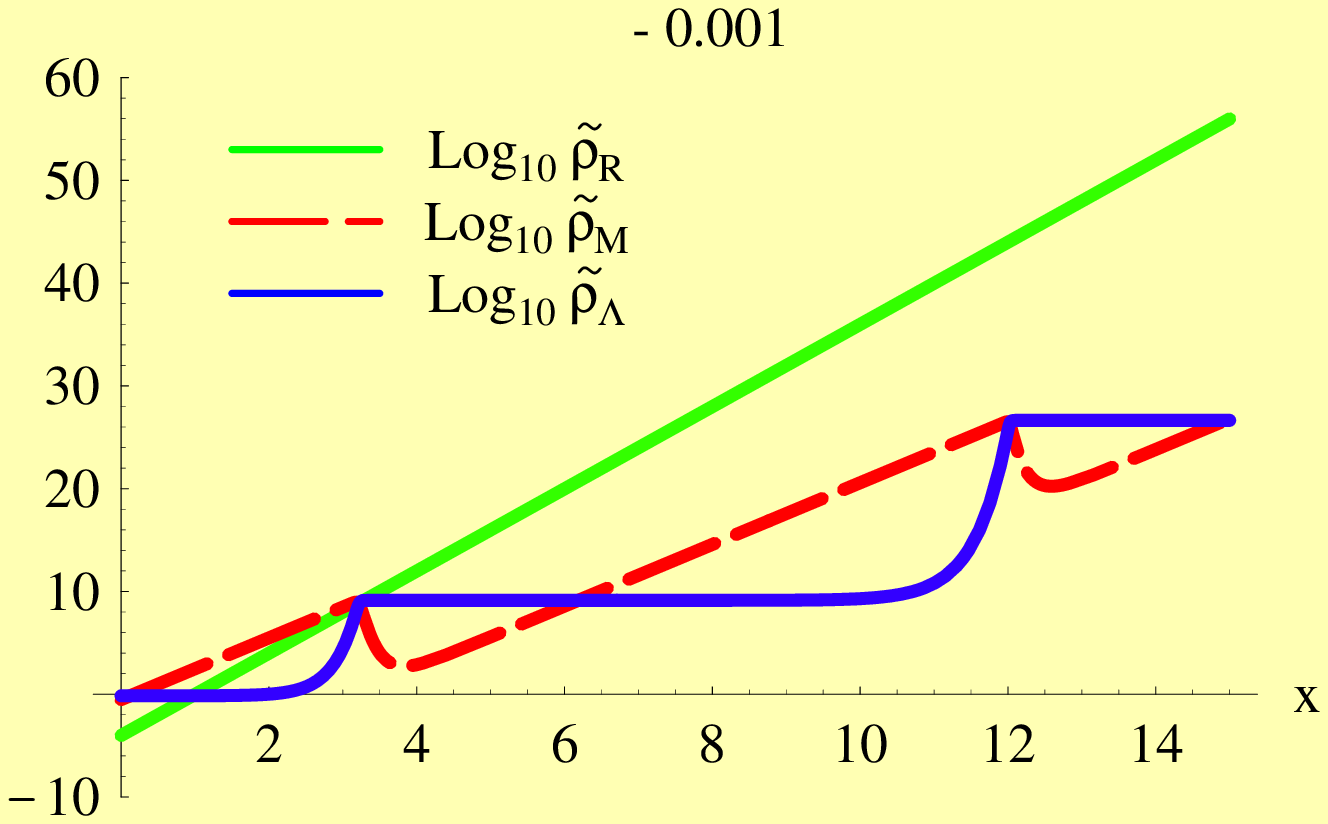}
\\
\epsfxsize=7cm \epsfysize=4.5cm \epsffile{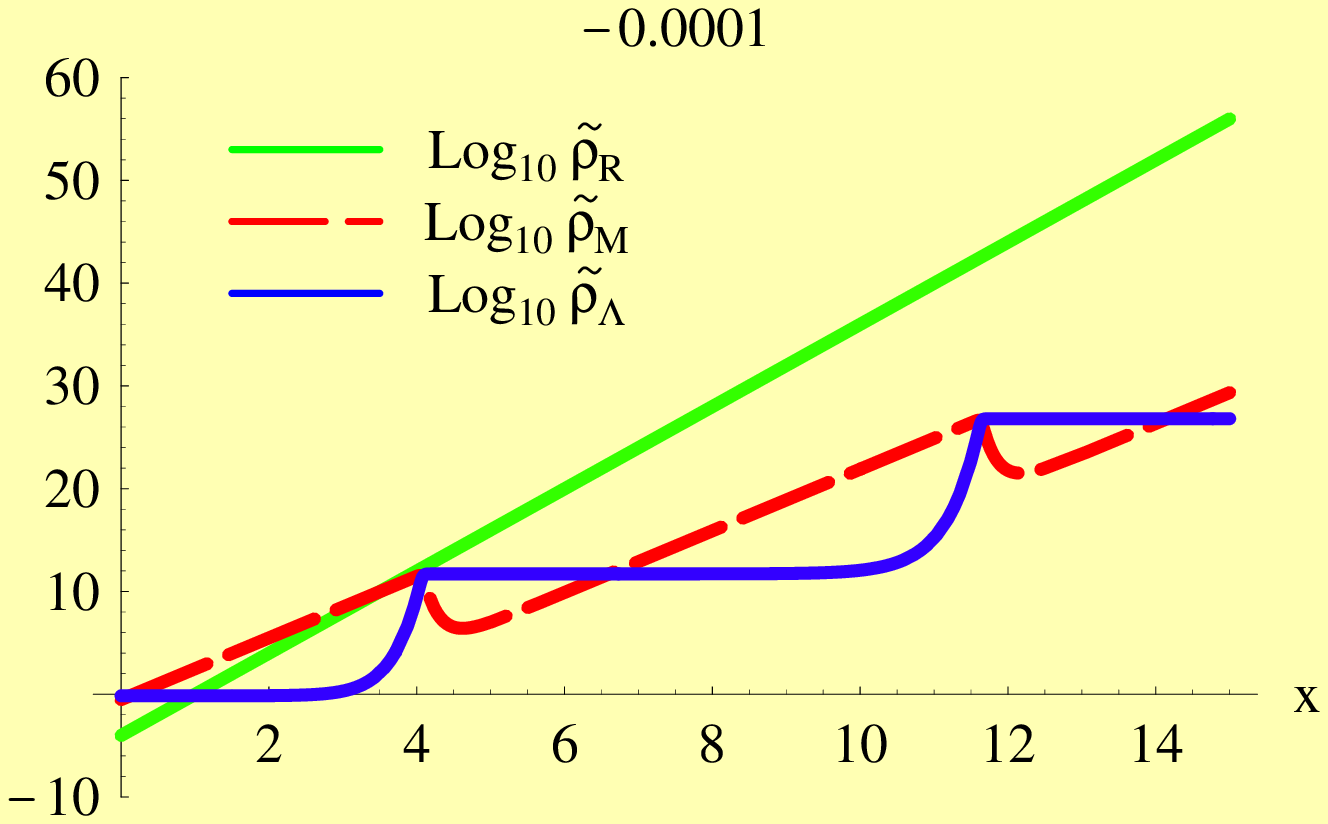} &
\epsfxsize=7cm \epsfysize=4.5cm \epsffile{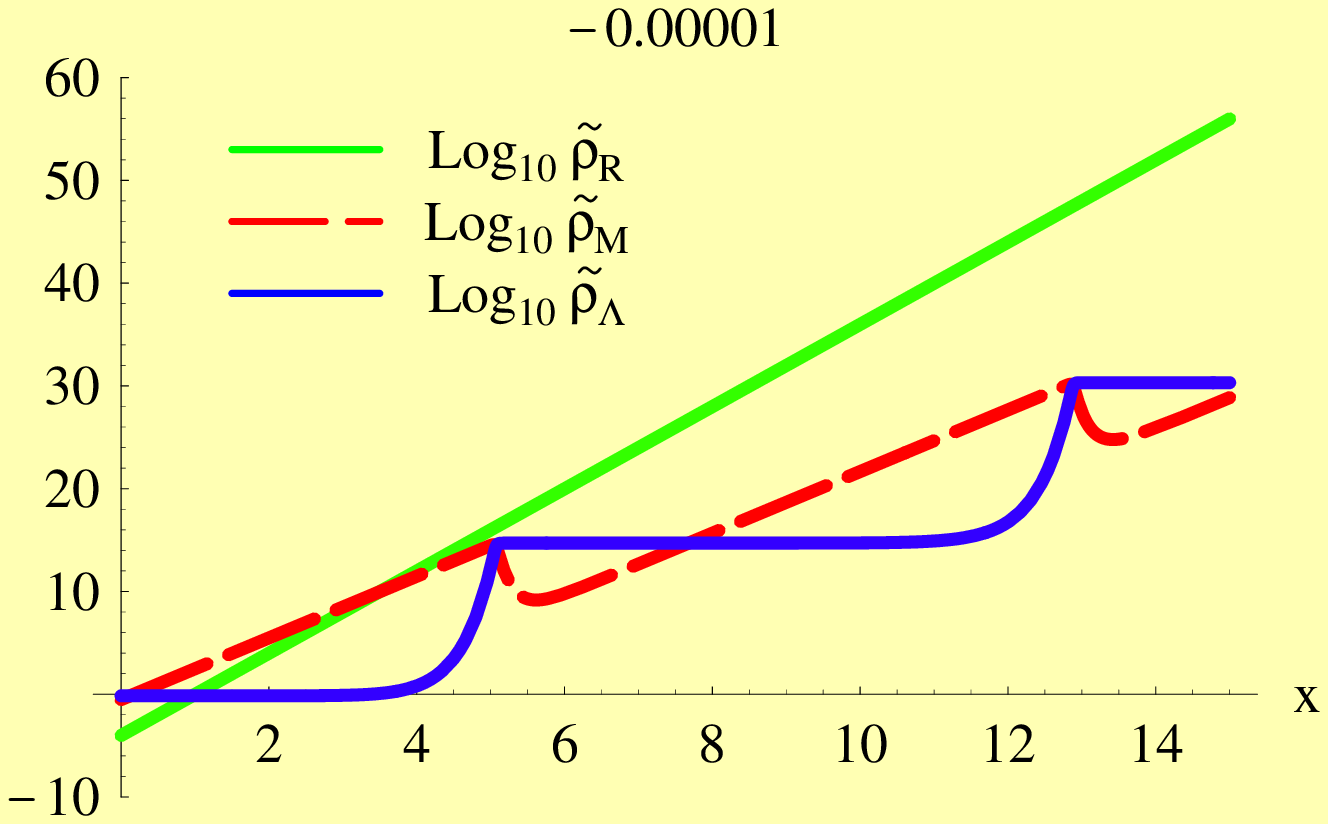}
\\
\epsfxsize=7cm \epsfysize=4.5cm \epsffile{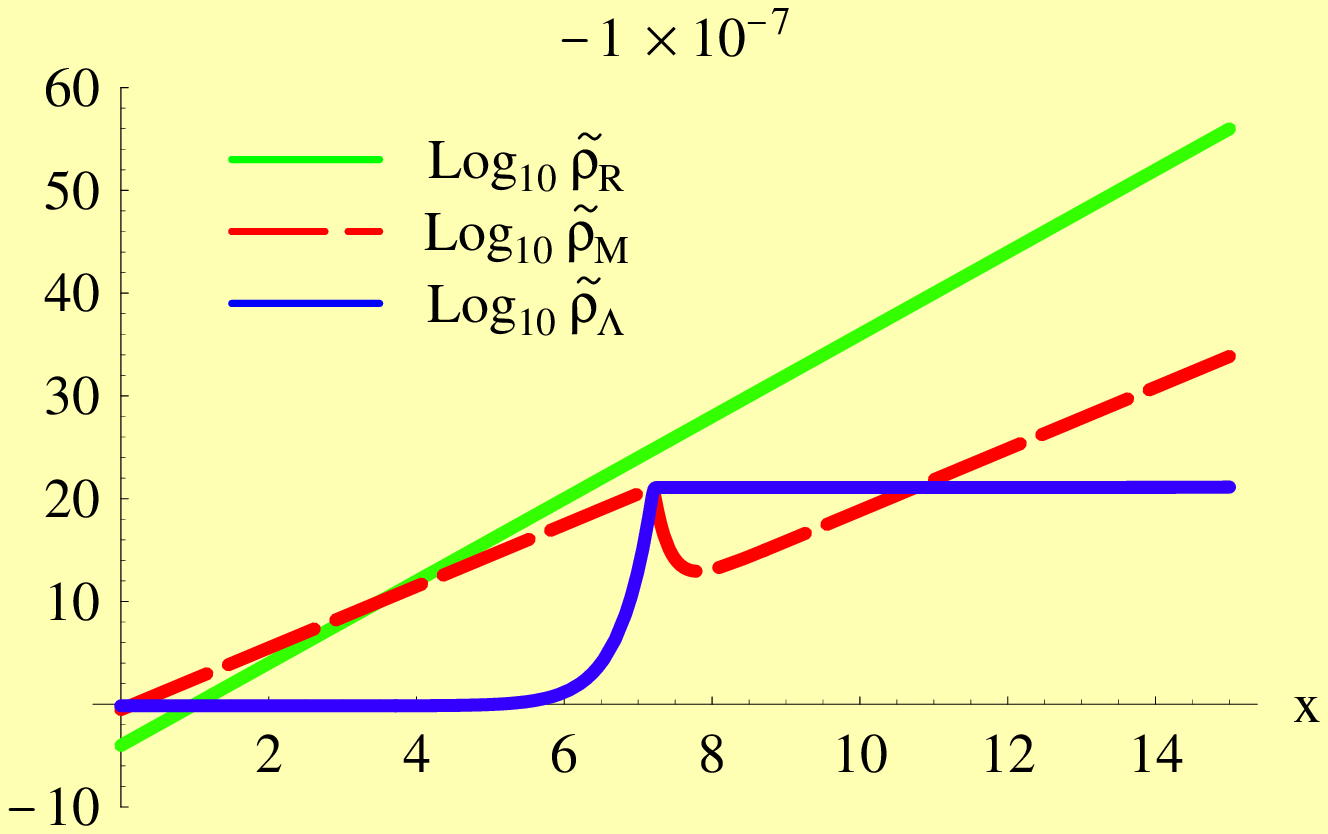} &
\epsfxsize=7cm \epsfysize=4.5cm
\epsffile{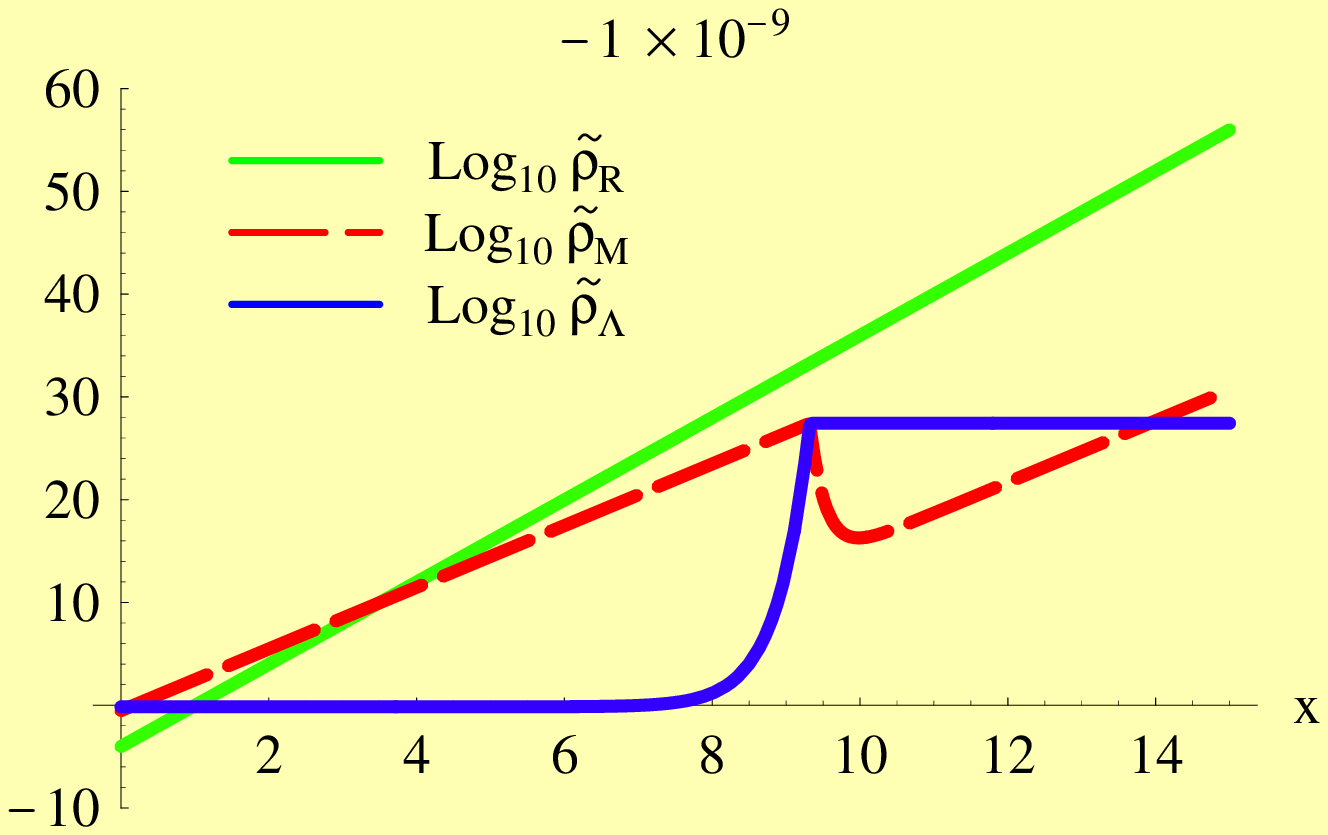}\\
\hline
\end{tabular}
\end{center}
\caption{Energy densities in unit of $\rho _{c_{0}}$ for varying
$\lambda$; the corresponding value of $\lambda$ is reported on the
top of each graph. The quantities  $\alpha$ and $\beta$ are both
fixed to be equal to $1$.} \label{varialambda}
\end{figure}

The peculiar features of the model do not critically depend on the value of
$\lambda$. In Figure \ref{varialambda} the energy density plots are shown
for several values of $\lambda$ in the range $ -1<\lambda <-10^{-9}$. As it
is clear from the figure, the redshift at the equivalence increases for
smaller value of $|\lambda|$. If one imposes the equivalence redshift in
the interval $z_{eq}=3100_{-400}^{+600}$ \cite{BowenMelchiorri} then
$\lambda$ must satisfy the bound $|\lambda |<10^{-4}$. Reminding the
definition of $\lambda$ (see Eq.(\ref{deflambda})), it is possible to
recast the bound on this parameter, for a given choice of $\alpha$ and
$\beta$, in a bound for the mass parameter $M$ and the dimensionless
coupling $k$ defined in Eq.(\ref{deflambda}). In particular for
$\alpha=\beta=1$ one would get
\begin{equation}
k \, \sqrt{\frac{3 \rho
_{c_{0}}}{8 \pi}} \, \frac{M_{Pl}}{M^3} \leq 10^{-4} \,\,\,\,\, \Longrightarrow
\,\,\,\,\, k \leq \frac{2.6}{h} \, \left( \frac{M}{GeV}\right)^3\, \,\,\,,
\label{bound}
\end{equation}
For values of the mass scale $M$ already larger than, say, few $GeV$,
Eq.(\ref{bound}) is not constraining at all the coupling $k$.

\begin{figure}
\begin{center}
\epsfxsize=13cm \epsfysize=8cm \epsffile{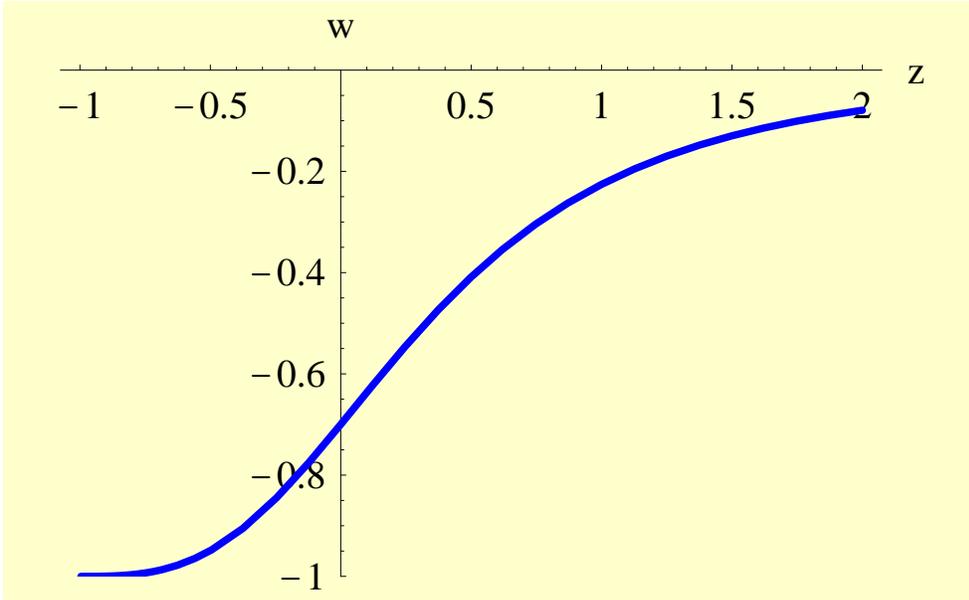}
\end{center}
\caption{Total equation of state for $\lambda =-0.001$ and $\alpha
=\beta =1$. The behaviour of $w$ is plotted for a range of $z$
near the present epoch.} \label{figwintorno}
\end{figure}

\begin{figure}
\begin{center}
\epsfxsize=13cm \epsfysize=8cm
\epsffile{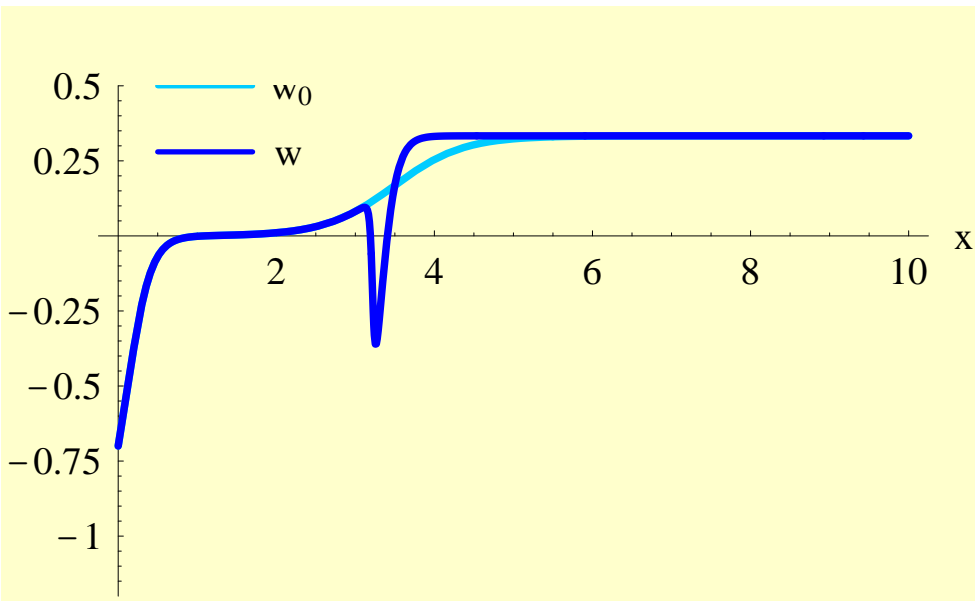}
\end{center}
\caption{Total equation of state for $\lambda =0$ ($w_0$) and
$\lambda=-0.001 ($w$)$ with $\alpha =\beta =1$.} \label{w}
\end{figure}

\begin{figure}
\begin{center}
\begin{tabular}{|cc|}
\hline &
\\
\epsfxsize=7cm \epsfysize=4.5cm \epsffile{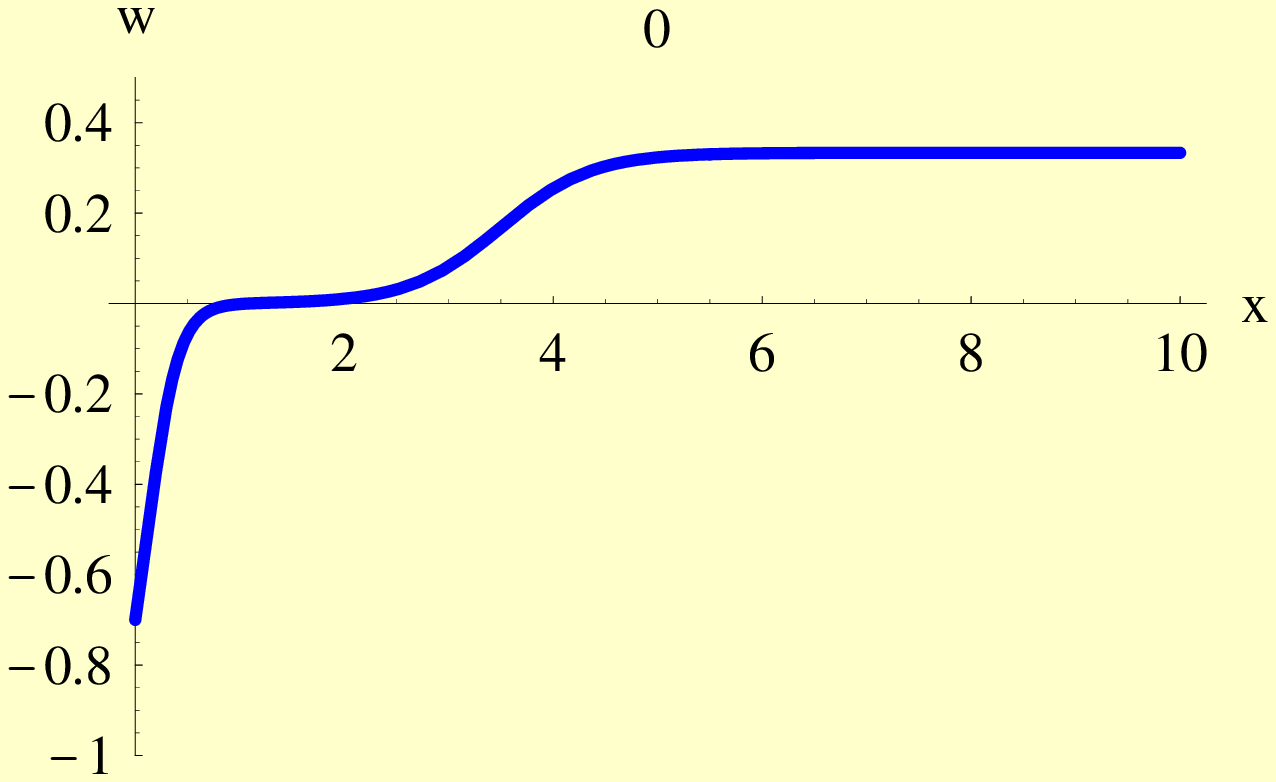} &
\epsfxsize=7cm \epsfysize=4.5cm \epsffile{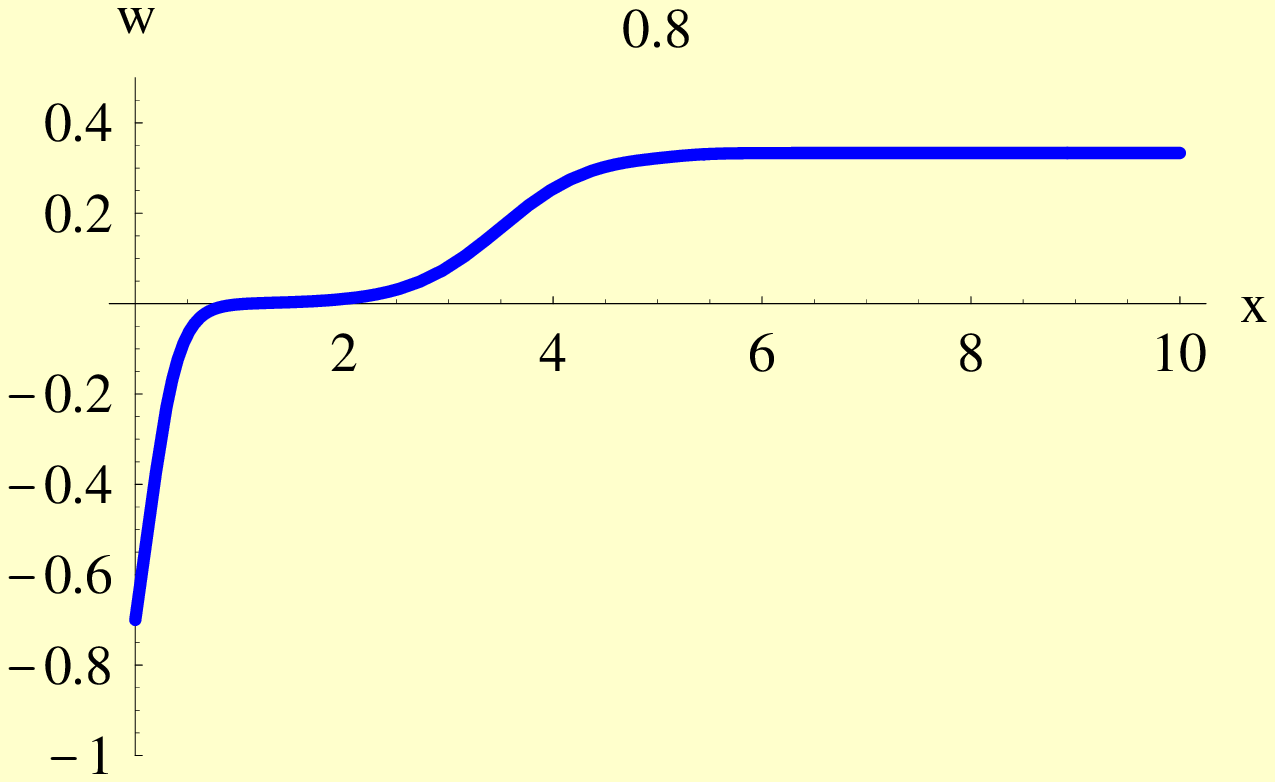}
\\
\epsfxsize=7cm \epsfysize=4.5cm \epsffile{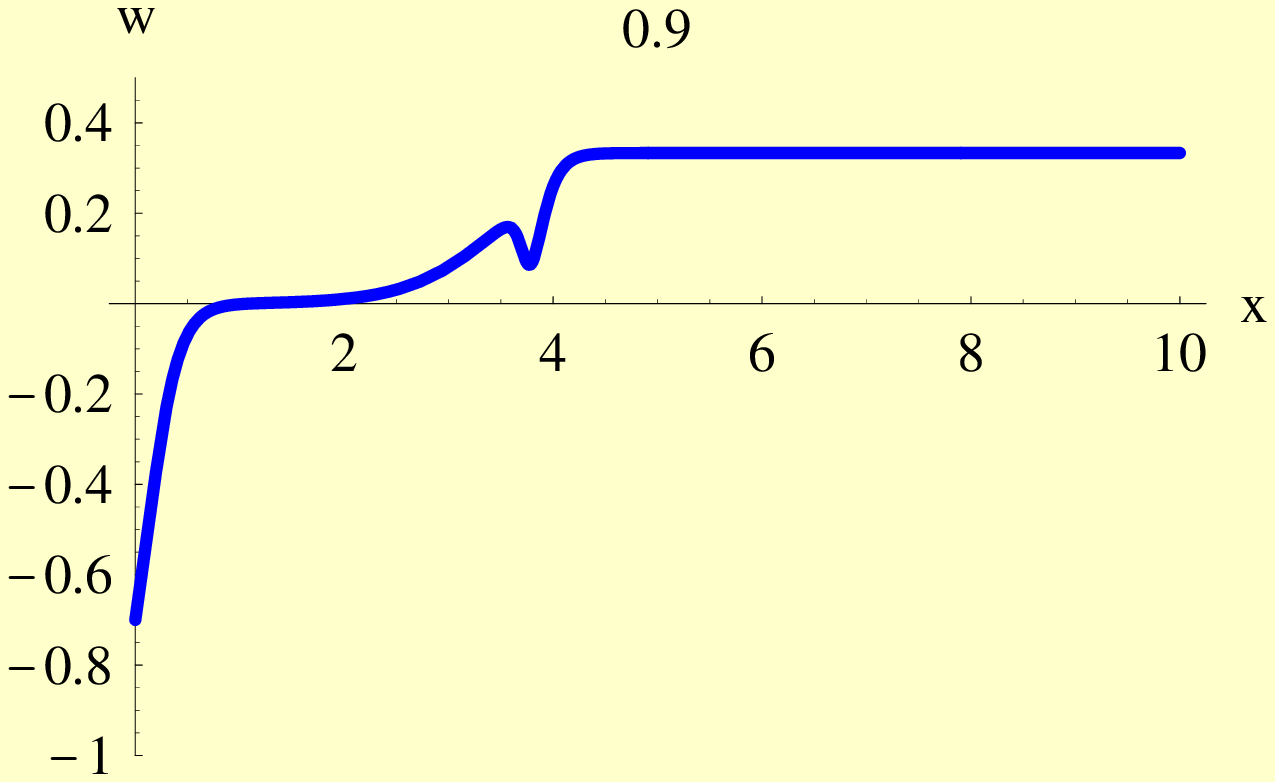} &
\epsfxsize=7cm \epsfysize=4.5cm \epsffile{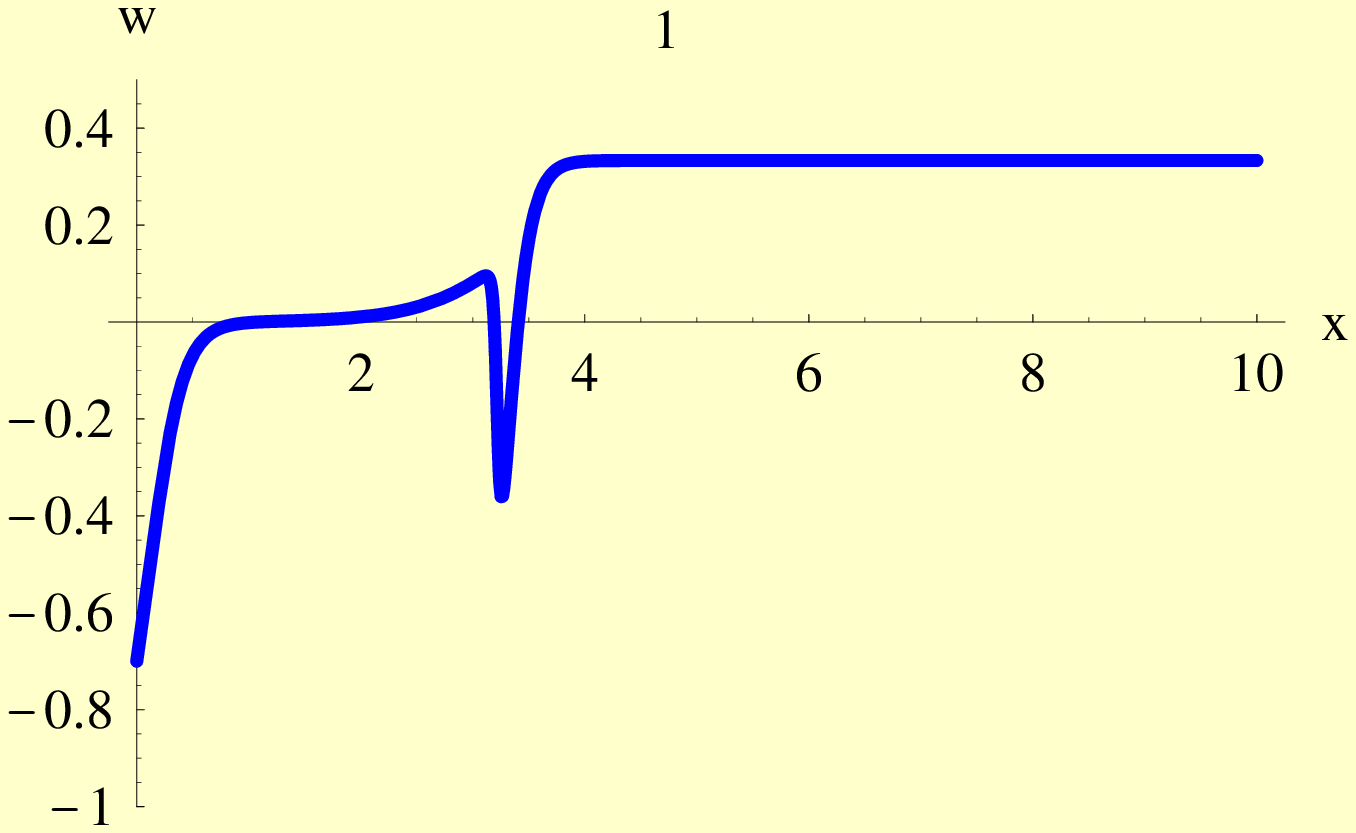}
\\
\hline
\end{tabular}
\end{center}
\caption{Total equation of state for different values of $\beta$,
with $\alpha=1$ and $\lambda=-0.001$.} \label{figwvariabeta}
\end{figure}

\begin{figure}
\begin{center}
\begin{tabular}{|cc|}
\hline &
\\
\epsfxsize=7cm \epsfysize=4.5cm \epsffile{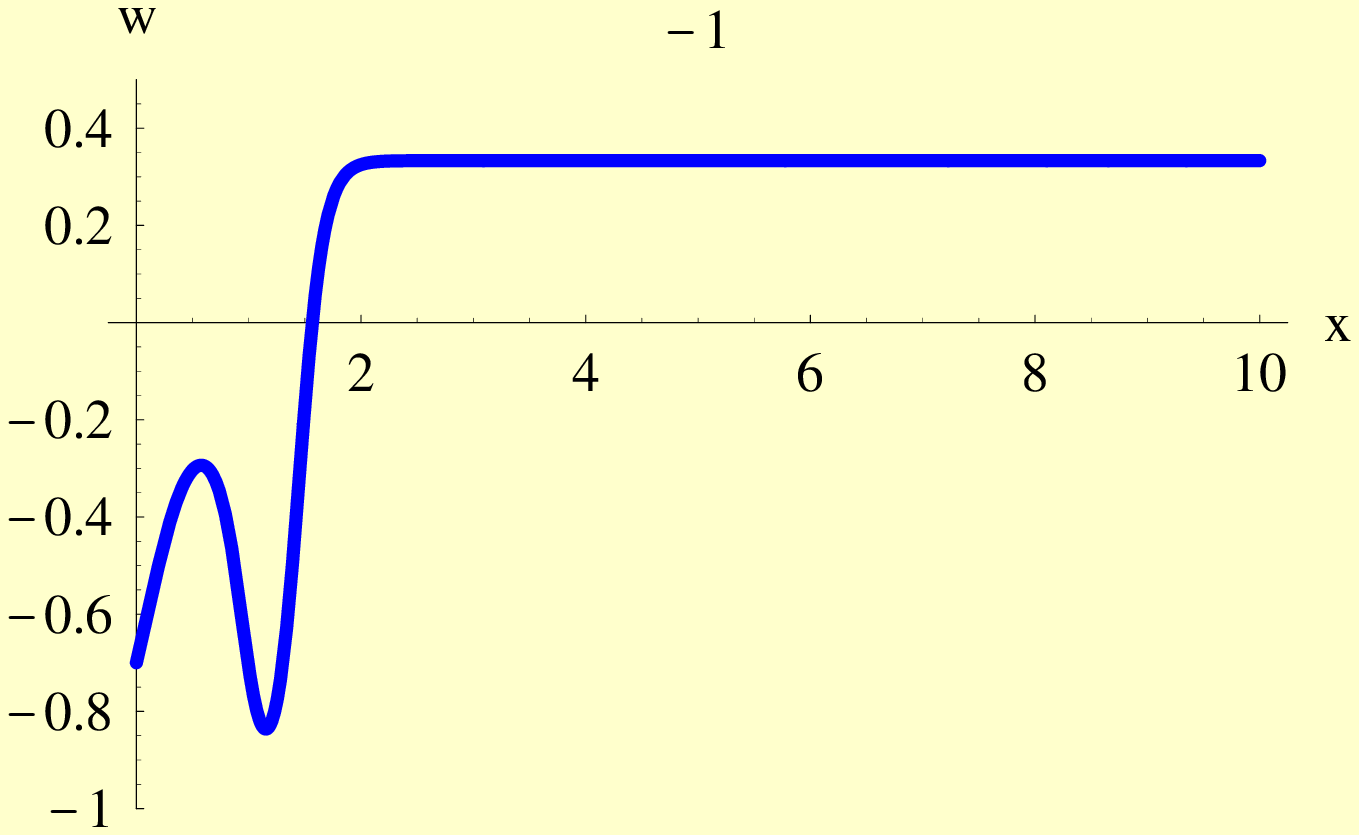} &
\epsfxsize=7cm \epsfysize=4.5cm \epsffile{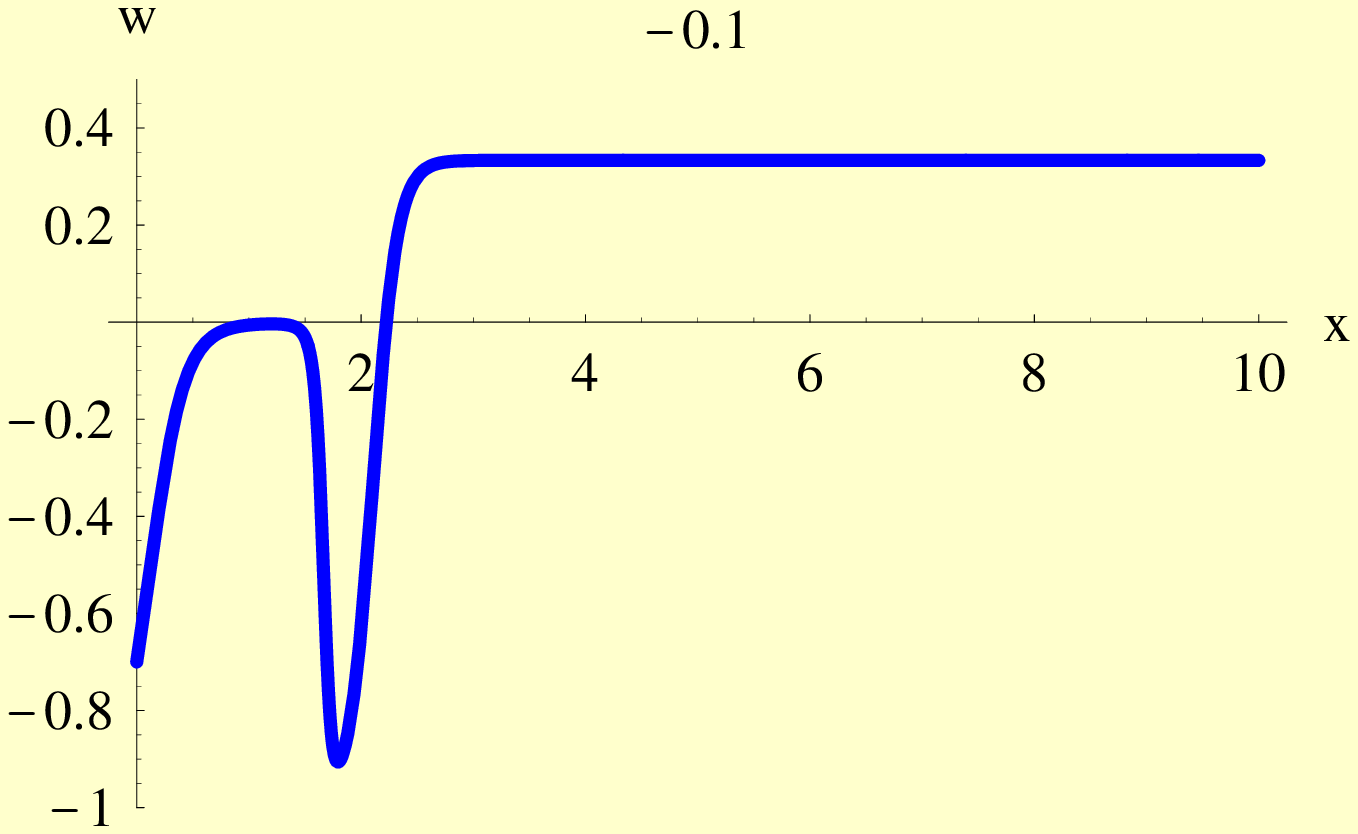}
\\
\epsfxsize=7cm \epsfysize=4.5cm \epsffile{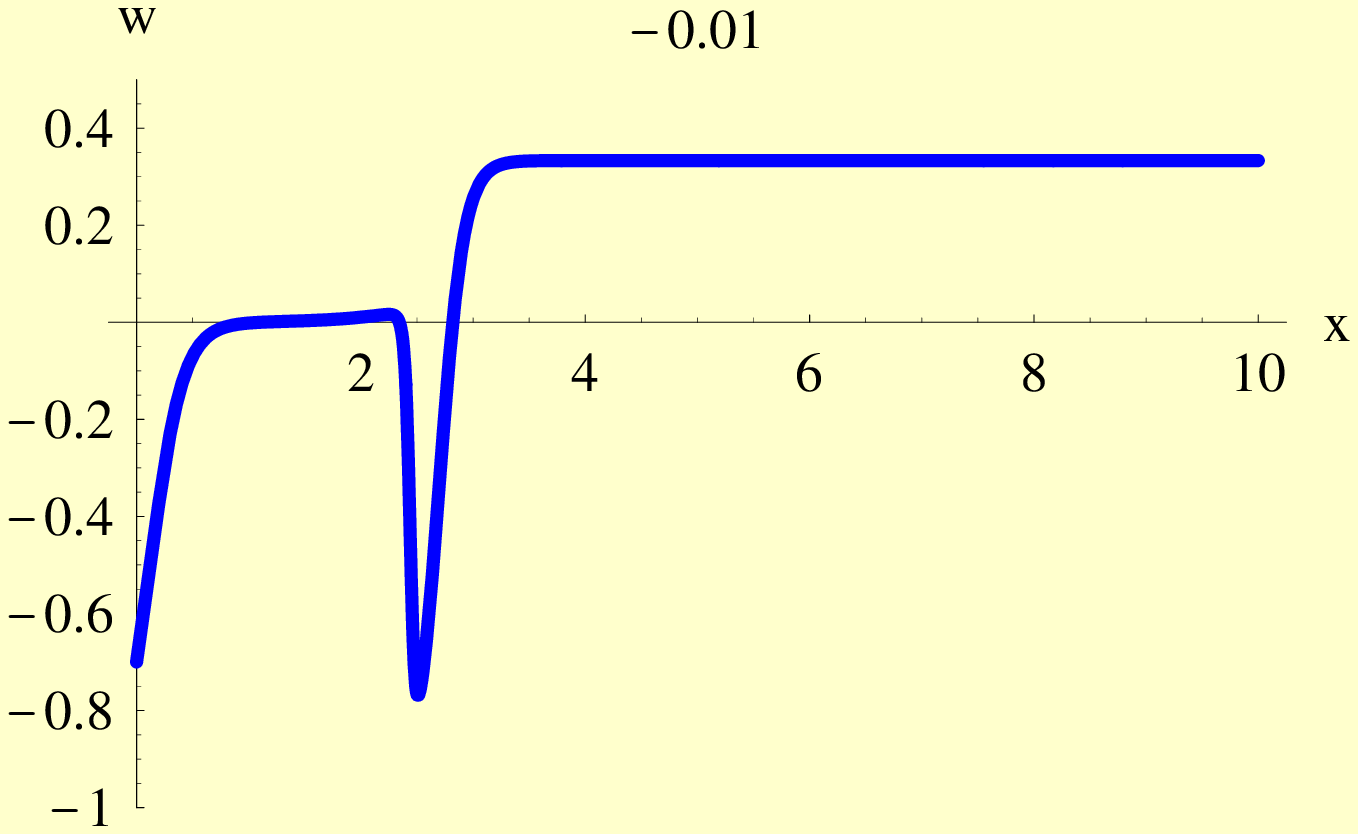} &
\epsfxsize=7cm \epsfysize=4.5cm \epsffile{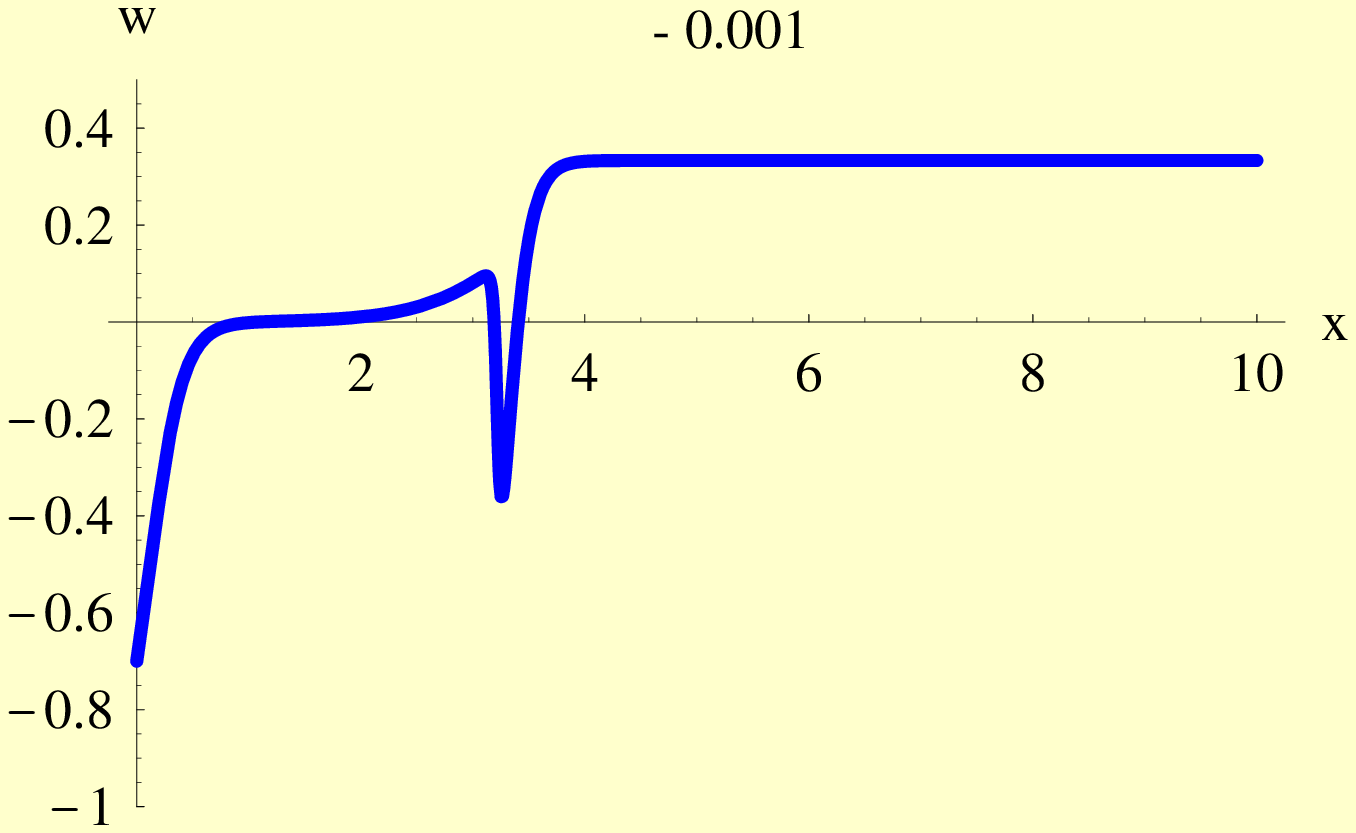}
\\
\epsfxsize=7cm \epsfysize=4.5cm \epsffile{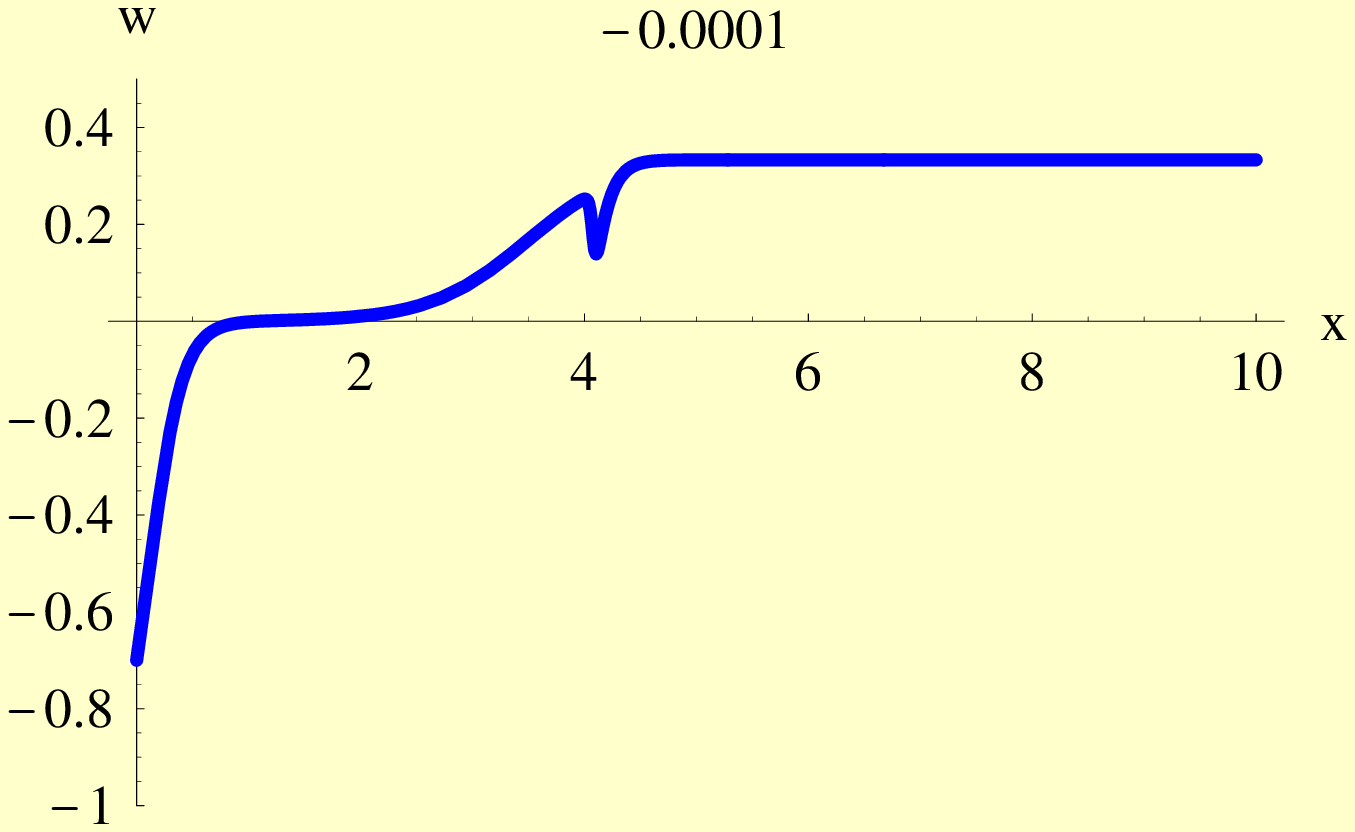} &
\epsfxsize=7cm \epsfysize=4.5cm \epsffile{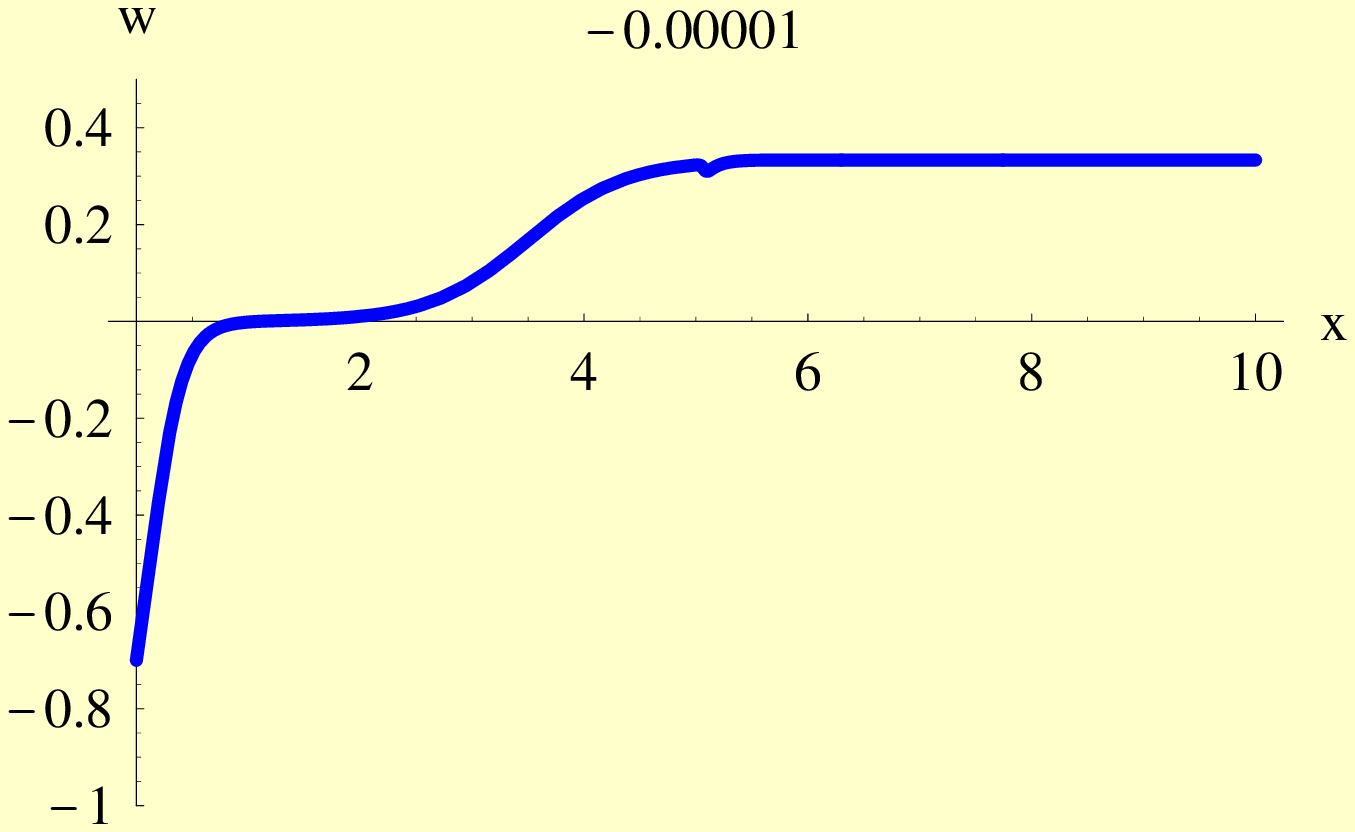}
\\
\hline
\end{tabular}
\end{center}
\caption{Total equation of state for different values of
$\lambda$, for $\alpha=\beta=1$.} \label{figwvarialambda}
\end{figure}

The main result of SN Ia data is that expansion is accelerated for $z
\simeq 1$. In order to understand if our model is indeed compatible with
this result we consider the effective equation of state of the primordial
fluid $w(x)$, defined as
\begin{equation}
w(x)\equiv \frac{\tilde{p}_{r}(x)+\tilde{p }_{m}(x)+\tilde{p
}_{\Lambda}(x)}{\tilde{\rho}_{r}(x)+\tilde{\rho }_{m}(x)+\tilde{\rho
}_{\Lambda}(x)}
=\frac{\frac{\tilde{\rho
}_{R}(x)}{3}-\tilde{\rho }_{\Lambda }(x)}{\tilde{\rho }_{M}(x) +\tilde{\rho
}_{R}(x)+\tilde{\rho }_{\Lambda }(x)}\,\,\,\,.
\end{equation}
In particular a phase of accelerated expansion is guaranteed if $w<-1/3$.
It is also worth noticing that recent data on the supernova SN1997ff at
$z=1.7$ indicates, moreover, that the universe expansion for this redshift
is still decelerated. This is actually in beautiful agreement with our
results, as can be seen in Figure \ref{figwintorno}, at least for this
choice of parameters. Figures \ref{figwintorno}, and \ref{figwvariabeta}
and \ref{figwvarialambda} also shows that the accelerated phase show that
accelaration occurs for $z \leq 1$ without assuming any particular
fine-tuning of free parameters. The behavior of  $w(x)$  for larger
redshifts and for two different values of $\lambda$ is reported in Figure
\ref{w}. The deep occurring for $x\sim 3.5$ is a genuine effect of the
interaction terms, which is more evident for large values of $\beta$ and/or
of $\lambda$ as shown in Figures \ref{figwvariabeta} and
\ref{figwvarialambda}, respectively. Note that the deep appears for
$|\lambda| \geq 10^{-5}$, and with the increasing of $|\lambda|$ moves to
smaller values of $x$.

Concerning the prediction for the age of the universe, as function of
$\tilde{\rho}_\Lambda$ and $\tilde{\rho}_m$, this does not provide any
severe bound on the free parameter ranges. For all choices of the
parameters which satisfy $z_{eq}=3100_{-400}^{+600}$ we have checked that
the result is the same, in the present uncertainty range, of a customary
non-interacting cosmological constant. This is simply due to the fact that,
during the last few red-shifts, which mainly contribute to the total age of
the universe, $\rho_\Lambda$ and $\rho_m$ effectively behave as a
cosmological constant and non interacting cold matter.

\begin{figure}
\begin{center}
\begin{tabular}{|cc|}
\hline &
\\
\epsfxsize=7cm \epsfysize=4.5cm \epsffile{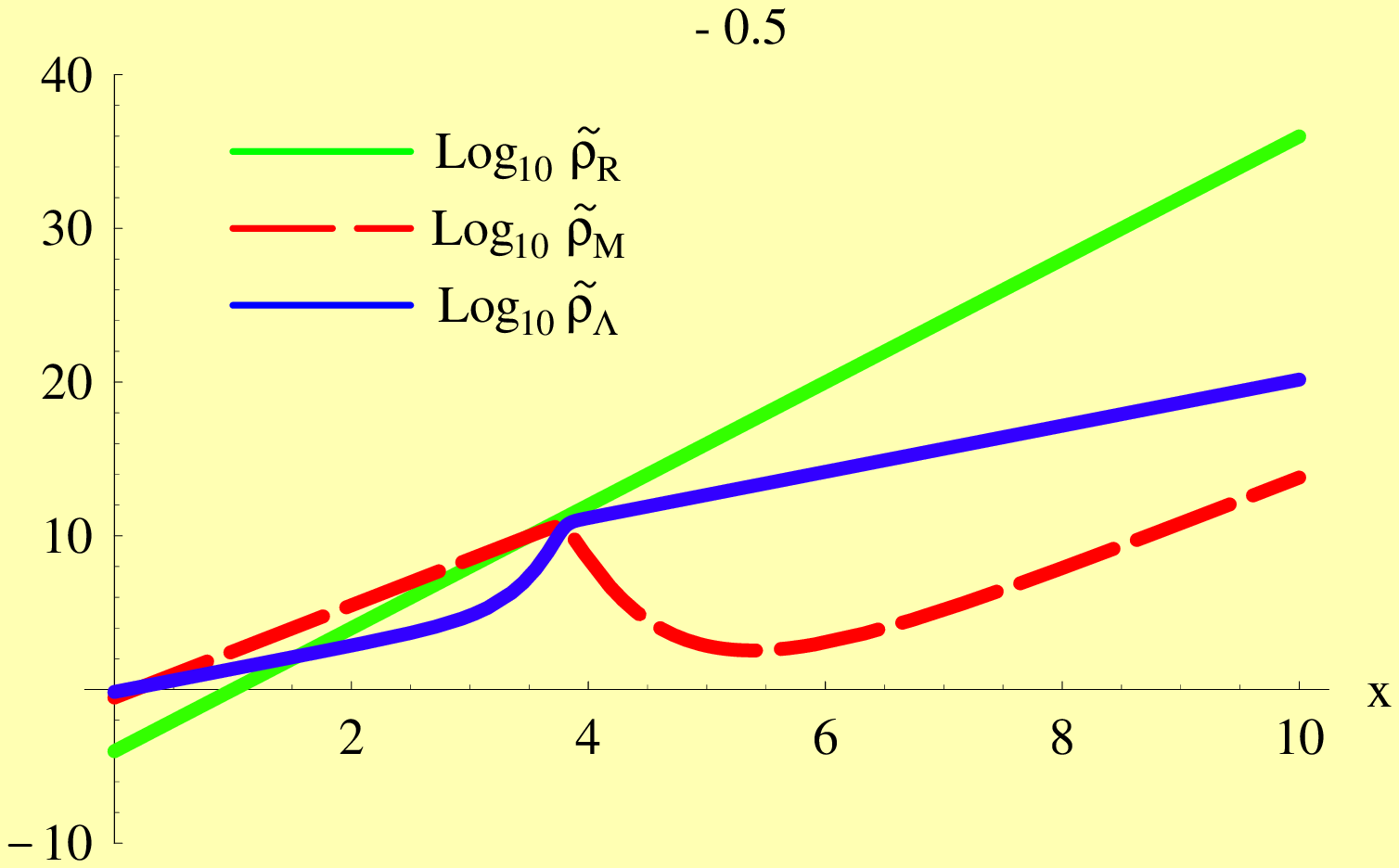} &
\epsfxsize=7cm \epsfysize=4.5cm \epsffile{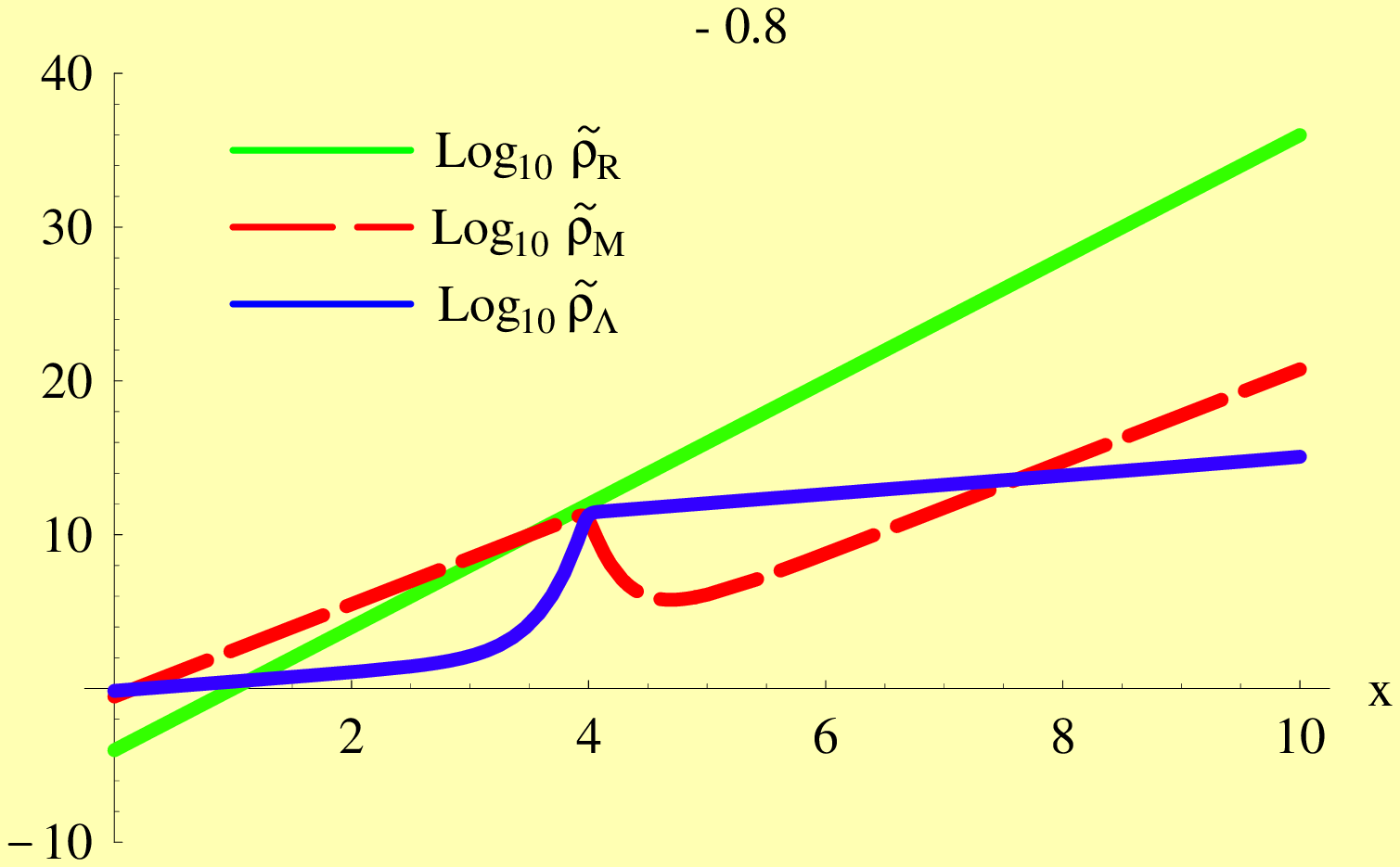}
\\
\hline
\end{tabular}
\end{center}
\caption{Energy densities in unit of $\rho _{c_{0}}$
for $\alpha=\beta=1$, $\lambda=-0.001$ and $\omega_\Lambda=-0.5$
and $-0.8$ respectively.} \label{fignew}
\end{figure}

The choice $\omega_\Lambda = -1$, even if affects of course the time
evolution of dark matter and dark energy, is not crucial to get the main
features of $\tilde{\rho}_\Lambda$ and $\tilde{\rho}_m$ evolution. In
particular a different choice for $\omega_\Lambda$, provided that $\omega
\leq -0.4$ is still producing the behaviour already discussed , with
$\tilde{\rho}_\Lambda$ and $\tilde{\rho}_m$ following and intersecting each
other during the universe evolution. This is shown in Figure \ref{fignew},
where the energy densities versus $x$ are plotted for $\alpha=\beta=1$,
$\lambda=-0.001$ and $\omega_\Lambda=-0.5$ and $-0.8$ respectively.

\begin{figure}
\begin{center}
\epsfxsize=13cm \epsfysize=8cm \epsffile{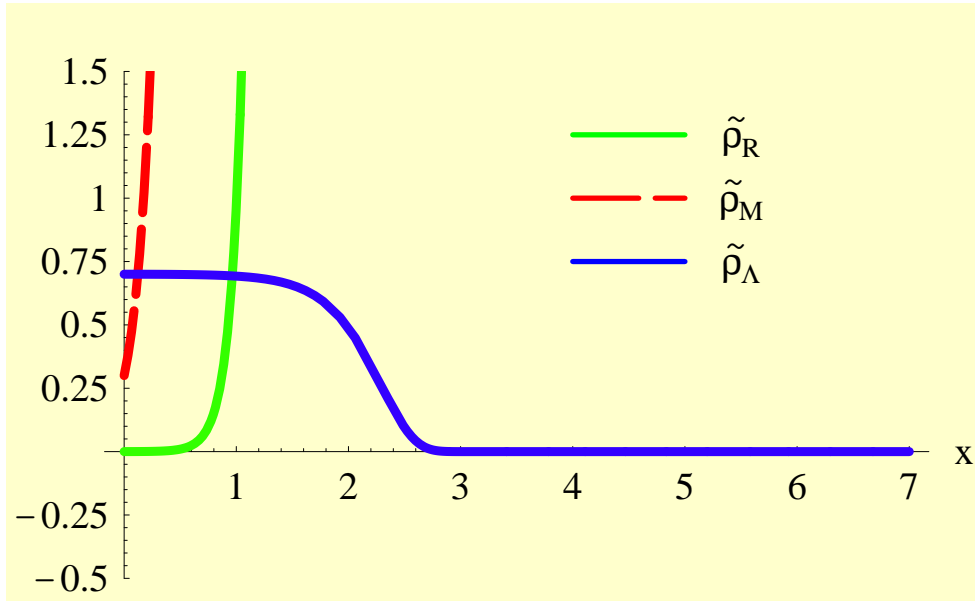}
\end{center}
\caption{Energy densities in critical units for $\lambda=0.0014$
and $\alpha=\beta=1$. } \label{figlambdaposnolog}
\end{figure}

As stated above the class of models with $\lambda
>0$ also describe a possibly interesting scenario. In this case, in fact,
the condensate is not produced till the redshift (temperature) does not
goes under a critical value, $x_c$. This is shown in Figure
\ref{figlambdaposnolog} where for the corresponding choice of parameters
$x_{c}\sim 2.8$. For $x$ larger than $x_c$, $\tilde{\rho}_\Lambda$ is
vanishing, and $\tilde{\rho}_m$ and $\tilde{\rho}_r$ have the same behavior
they would have in absence of interaction. At $x=x_c$ the fluid undergoes a
phase transition and part of the energy density stored in dark matter flows
in the condensate till it start dominate at low red-shifts. Unfortunately
this scenario is still affected by the coincidence problem. The value of
$x_c$ where the phase transition takes place is determined by the values of
$\rho_m$ and $\rho_\Lambda$ today or, viceversa, it should be accurately
tuned to provide the observed values for the matter/cosmological constant
energy ratio today.

To summarize, the coupled quintessence framework provides an interesting
scenario where the mutual interaction between dark matter and dark energy
components shows a quite close evolution of their energy densities during
the universe expansion. This remarkable feature represents a simple way out
to the coincidence problem, since it makes the present situation of
$\rho_\Lambda \sim \rho_m$ a cyclic condition which have occurred many
times in the past. The detail of the energy density evolution is of course
a function of the assumed interaction term. We have considered a simple
model where the dark energy is provided by a condensate component of a
fluid. The interaction with the gas of non relativistic particle (dark
matter) is then due to {\it stimulated} decay processes of the condensate
into gas particles. Though the model is still lacking a detailed
microscopic description, it is nevertheless quite phenomenologically well
motivated, and it would be worth analyzing their properties in more
details, in view of the natural way it deals with the cosmic coincidence
problem.

\noindent
{\bf Acknowledgements}\\ We would like to thank S. Matarrese for useful
discussions.

\begin{figure}
\begin{center}
\epsfxsize=13cm \epsfysize=8cm \epsffile{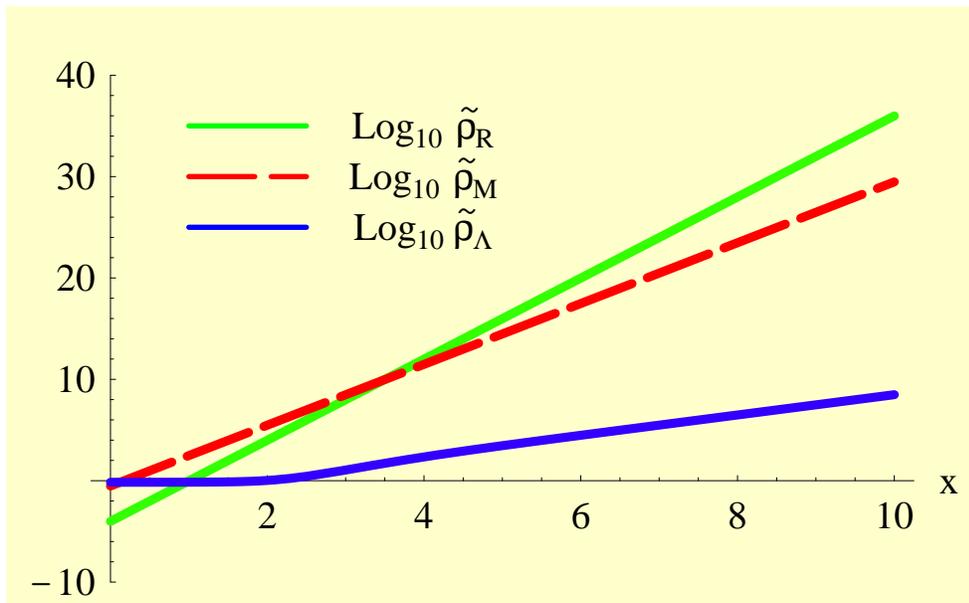}
\end{center}
\caption{Energy densities in unit of $\rho _{c_{0}}$ for $\alpha
=1$, $\beta =0$ and $\lambda =-0.001$.} \label{densita`
alfa1beta0}
\end{figure}

\end{document}